\documentclass[fleqn,usenatbib,useAMS]{mnras}

\usepackage{graphicx}	% Including figure files
\usepackage{amsmath}	% Advanced maths commands
\usepackage{amssymb}	% Extra maths symbols
\usepackage{multicol}        % Multi-column entries in tables
\usepackage{bm}		% Bold maths symbols, including upright Greek
\usepackage{pdflscape}	% Landscape pages

\usepackage[T1]{fontenc}
\usepackage{ae,aecompl}

\usepackage{newtxtext,newtxmath}
\newcommand{\warwick}{1}
\newcommand{\unam}{2}
\newcommand{\valpa}{3}
\newcommand{\naresuam}{4}
\newcommand{\maxplanck}{5}
\newcommand{\ljmu}{6}
\newcommand{\saao}{7}
\newcommand{\capet}{8}
\newcommand{\ufs}{9}
\newcommand{\vvs}{10}
\newcommand{\bav}{11}
\newcommand{\aavso}{12}
\newcommand{\uj}{13}
\newcommand{\hamburg}{14}
\newcommand{\sheffield}{15}
\newcommand{\iac}{16}
\newcommand{\cork}{17}
\newcommand{\ufrgs}{18}

\title[A targeted search for binary white dwarf pulsars]{A targeted search for binary white dwarf pulsars using {\it Gaia} and WISE}

\author[Pelisoli et al.]{
Ingrid Pelisoli$^{\warwick}$\thanks{E-mail: ingrid.pelisoli@warwick.ac.uk},
T.~R. Marsh$^{\warwick}$,
G.~Tovmassian$^{\unam}$, %PI of dedicated proposal
L.~A. Amaral$^{\valpa}$,
Amornrat Aungwerojwit$^{\naresuam}$,
M.~J. Green$^{\maxplanck}$,\newauthor
R.~P. Ashley$^{\ljmu}$,
David A.~H. Buckley$^{\saao,\capet,\ufs}$,
B.~T. G\"ansicke$^{\warwick}$,
F.-J. Hambsch$^{\vvs,\bav, \aavso}$,
K. Inight$^{\warwick}$,
S.~B. Potter$^{\saao, \uj}$,\newauthor
A.~J. Brown$^{\hamburg}$,
N. Castro Segura$^{\warwick}$,
V.~S. Dhillon$^{\sheffield,\iac}$,
M.~J. Dyer$^{\sheffield}$,
J.~A. Garbutt$^{\sheffield}$,
D. Jarvis$^{\sheffield}$,\newauthor
M. R. Kennedy$^{\cork}$,
S.~O. Kepler$^{\ufrgs}$,
P. Kerry$^{\sheffield}$,
S.~P. Littlefair$^{\sheffield}$,
J. McCormac$^{\warwick}$,\newauthor
J. Munday$^{\warwick}$,
S.~G. Parsons$^{\sheffield}$,
E. Pike$^{\sheffield}$,
D.~I. Sahman$^{\sheffield}$
\\
$^{\warwick}$Department of Physics, University of Warwick, Gibbet Hill Road, Coventry, CV4 7AL, UK\\
$^{\unam}$Universidad Nacional Aut\'{o}noma de M\'{e}xico, Instituto de Astronom\'{i}a, Aptdo Postal 106, Ensenada 22860, Baja California, M\'{e}xico\\
$^{\valpa}$Instituto de F\'{i}sica y Astronom\'{i}a, Universidad de Valpara\'{i}so, Gran Breta\~{n}a 1111, Playa Ancha, Valpara\'{i}so 2360102, Chile\\
$^{\naresuam}$Department of Physics, Faculty of Science, Naresuan University, Phitsanulok, 65000, Thailand\\
$^{\maxplanck}$Max-Planck-Institut f\:{u}r Astronomie, K\:{o}nigstuhl 17, D-69117 Heidelberg, Germany\\
$^{\ljmu}$Astrophysics Research Institute, Liverpool John Moores University, 146 Brownlow Hill, L3 5RF, Merseyside, UK\\
$^{\saao}$South African Astronomical Observatory, PO Box 9, Observatory, 7935, Cape Town, South Africa\\
$^{\capet}$Department of Astronomy, University of Cape Town, Private Bag X3, Rondebosch 7701, South Africa\\
$^{\ufs}$Department of Physics, University of the Free State, PO Box 339, Bloemfontein 9300, South Africa\\
$^{\vvs}$Vereniging Voor Sterrenkunde (VVS), Zeeweg 96, 8200 Brugge, Belgium\\
$^{\bav}$Bundesdeutsche Arbeitsgemeinschaft fur Ver\"{a}nderliche Sterne (BAV), Munsterdamm 90, 12169 Berlin, Germany\\
$^{\aavso}$American Association of Variable Star Observers, Cambridge, MA 02138, USA\\
$^{\uj}${Department of Physics, University of Johannesburg, PO Box 524, Auckland Park 2006, South Africa}\\
$^{\hamburg}$Hamburger Sternwarte, University of Hamburg, Gojenbergsweg 112, 21029 Hamburg, Germany\\
$^{\sheffield}$Astrophysics Research Cluster, School of Mathematical and Physical Sciences, University of Sheffield, Sheffield, S3 7RH, United Kingdom\\
$^{\iac}$Instituto de Astrof\'{i}sica de Canarias, E-38205 La Laguna, Tenerife, Spain\\
$^{\cork}$School of Physics, University College Cork, Cork, T12 K8AF, Ireland\\
$^{\ufrgs}$Instituto de F\'{i}sica, Universidade Federal do Rio Grande do Sul, 91501-970 Porto Alegre, RS, Brazil\\
}

\date{Last updated XXX; in original form XX}

\pubyear{2025}

\begin{document}
\label{firstpage}
\pagerange{\pageref{firstpage}--\pageref{lastpage}}
\maketitle

% Abstract of the paper
\begin{abstract}
After its discovery in 2016, the white dwarf binary AR~Scorpii (AR Sco) remained for several years the only white dwarf system to show pulsed radio emission associated with a fast-spinning white dwarf. The evolutionary origin and the emission mechanism for AR~Sco are not completely understood, with different models proposed. Testing and improving these models requires observational input. Here we report the results of a targeted search for other binary white dwarf pulsars like AR~Sco. Using data from {\it Gaia} and WISE, we identified 56 candidate systems with similar properties to AR~Sco, of which 26 were previously uncharacterised. These were subject to spectroscopic and photometric follow-up observations. Aside from one new binary white dwarf pulsar found, J191213.72-441045.1, which was reported in a separate work, we find no other systems whose characteristics are akin to AR~Sco. The newly characterised systems are primarily young stellar objects (with 10 found) or cataclysmic variables (7 identifications), with the remaining being either blended or non-variable on short timescales.
\end{abstract}

% Select between one and six entries from the list of approved keywords.
% Don't make up new ones.

%%%%%%%%%%%%%%%%%%%%%%%%%%%%%%%%%%%%%%%%%%%%%%%%%%

%%%%%%%%%%%%%%%%% BODY OF PAPER %%%%%%%%%%%%%%%%%%

\section{Introduction}

Non-thermal emission from stars is a sign of relativistic particle acceleration in the presence of magnetic fields. Most commonly detected from pulsars and black-hole jet sources, the discovery of AR~Scorpii (henceforth AR~Sco) showed that white dwarfs can also generate substantial levels of non-thermal flux \citep{Marsh2016}. AR~Sco is a binary system composed of a rapidly-spinning, magnetic white dwarf and a low-mass main sequence star of M-type. Its most striking characteristic is pulsed emission that is detected from radio \citep{Stanway2018} to X-rays \citep{Takata2018} on a period of 1.97~minutes. Although radio emission has been detected in accreting white dwarfs \citep[e.g.][]{Coppejans2015, Barrett2020, Ridder2023} and from the two so-called magnetic propellers \citep{Bookbinder1987, Pretorius2021}, where material is being transferred but is ejected before accretion onto the white dwarf, AR~Sco was the first white dwarf showing radio pulses in the absence of mass transfer, which led to it becoming known as a white dwarf pulsar.

Similar to neutron star pulsars, the pulsed luminosity of AR~Sco is powered by the rapid spin-down of the white dwarf \citep[$P/\dot{P} \sim 5 \times 10^{6}$~year,][]{Stiller2018, Gaibor2020, Pelisoli2022c}, and its broad-band spectral energy distribution is consistent with synchrotron emission from relativistic electrons. However, in the case of AR~Sco, binary interaction seems to be the main driver of the emission\footnote{For this reason, we refer to AR~Sco as a {\it binary} white dwarf pulsar throughout this work.}. The emission is most likely triggered by the injection of particles into the magnetosphere of the white dwarf as it sweeps past the cool companion \citep{Geng2016, PotterBuckley2018}, but the exact mechanism behind the pulses is not fully understood. 

The rapid spin-down of AR~Sco suggests that it could be an evolutionary link between the two main classes of accreting magnetic white dwarfs, polars \citep[also known as AM Her stars,][]{Cropper1990} and intermediate polars \citep[IPs or DQ Her stars,][]{Patterson1994}. In polars, the white dwarf spin is synchronised with the orbital period of the binary, whereas in IPs the white dwarf spin period is faster than the orbital period. A model for this possible evolutionary link has been proposed by \citet{Schreiber2021}. They proposed that accreting magnetic white dwarfs are originally not strongly magnetic, which allows them to accrete efficiently from the companion, gaining angular momentum that causes spin-up. As the white dwarf cools, the core starts to crystallise, which, combined with the fast rotation, potentially creates the conditions for a dynamo \citep{Isern2017, Ginzburg2022}. Once a magnetic field emerges, the observed system is an accreting magnetic white dwarf spinning faster than the orbit -- an IP. When the white dwarf's magnetic field becomes strong enough, it will connect with the magnetic field of the M-dwarf, providing a synchronising torque that causes the white dwarf to spin-down and transfers angular momentum from its spin to the orbit. That may cause the system to become detached, and be observed as a binary white dwarf pulsar, with a fast spin-down and no accretion, like AR~Sco. Over time, the white dwarf spin will synchronise with the orbit, and angular momentum loss due to magnetic braking and gravitational wave radiation will bring the stars closer together again, mass transfer restarts, and the system thus becomes a polar.

The models for AR~Sco's pulsed emission, as well as the model for its formation, require observational input to be tested. In particular, there is still uncertainty as to whether a crystallisation dynamo is possible for white dwarfs \citep[e.g.][]{Fuentes2023}. One of the predictions of \citet{Schreiber2021}'s evolutionary model is that binary white dwarf pulsars are a possible evolutionary stage in the evolution of accreting white dwarfs, suggesting that binary white dwarf pulsars should be a class of systems. The discovery of J191213.72-441045.1 \citep[J1912-4410,][]{Pelisoli2023,Schwope2023} provides support for this hypothesis, although it is unclear whether J1912-4410 fits the crystallisation model given the white dwarf's estimated physical parameters \citep{Pelisoli2024a}. A potential connection between these systems and long-period radio transients (LPTs), some of which have been found to harbour white dwarf binaries \citep{deRuiter2024, HurleyWalker2024, Rodriguez2025}, is another avenue that requires further investigation that would benefit from a larger sample of binary white dwarf pulsars allowing a better picture of the characteristics of this class.

Arguably one of the reasons why so few binary white dwarf pulsars are known is the lack of systematic searches for similar systems. AR~Sco itself was mis-classified as a $\delta$-Scuti variable for decades \citep{Satyvaldiev1971} before its then unique nature was discovered by citizen astronomers. A possible exception is the recent search for radio emission from white dwarfs using data from the Very Large Array Sky Survey (VLASS) Epoch 1 Quick Look Catalogue \citep{Gordon+20, Gordon+21} carried out by \citet{Pelisoli2024b}. However, this search utilised a white dwarf catalogue \citep{GentileFusillo+21} whose selection criteria are tailored for isolated white dwarfs and in fact exclude AR~Sco. The search uncovered no secure radio detection from a white dwarf above the VLASS threshold (1--3~mJy), with only one white dwarf system possibly having associated radio emission, though other observational evidence favours a chance alignment. Their main conclusion was that strong radio emission from white dwarfs must be rare outside of interacting binaries, therefore an efficient selection for white dwarf pulsars needs to take binarity into account. In this work, we take that into consideration and report the results of a systematic search for binary white dwarf pulsars. We selected candidates showing similar observational properties to AR~Sco and carried out follow-up observations to characterise them.

\section{Candidate Selection}
\label{sec:selection}

\subsection{{\it Gaia} eDR3}

\begin{figure}
    \includegraphics[width=\columnwidth]{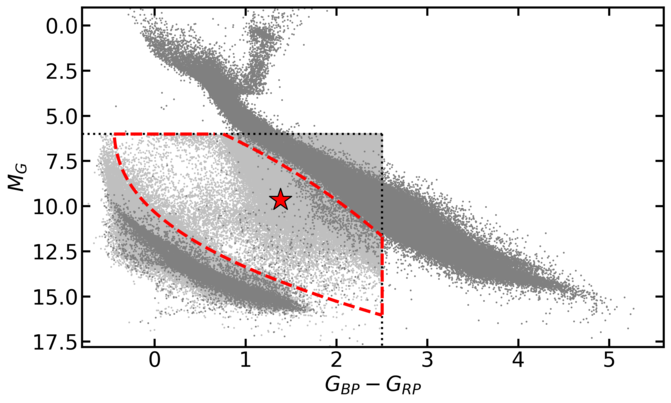}
    \caption{The first step in the candidate selection, using {\it Gaia} only. The red star shows the position of AR~Sco. Our initial selection is marked by the black dotted lines. The lower $M_G$ limit was defined to exclude hot subdwarf stars, and the upper $G_{BP} - G_{RP}$ limit aims at excluding low-mass main sequence stars and brown dwarfs. The objects initially selected are shown in light grey, whereas in dark grey we show the population within 100~pc to illustrate the location of main sequence and of the white dwarf cooling track. The red dashed lines show the final {\it Gaia} selection, excluding single white dwarfs and main sequence stars.}
    \label{fig:gaia}
\end{figure}

Given its nature as a binary system containing a white dwarf and a cool main sequence star, AR~Sco sits in a low-density region in the {\it Gaia} colour-magnitude diagram, as shown in Fig.~\ref{fig:gaia}. Therefore, the first step in our candidate selection was to identify sources in the same region of this diagram as AR~Sco. We initially selected objects in a broad region defined by:
\begin{eqnarray}
   G_{BP} - G_{RP} &<& 2.5\\
   M_G &>& 6
\end{eqnarray}
where $G_{BP}$ and $G_{RP}$ are the apparent magnitudes in the  {\it Gaia} ${BP}$ and ${RP}$ passbands, and $M_G$ is the absolute magnitude in the {\it Gaia} $G$ passband, initially calculated without any parallax zero-point correction, that is $5 + 5\log(\varpi/1000) + G$, where $\varpi$ is the reported parallax. It is worth noting that, even though our selection was applied using eDR3 data, astrometric and photometric measurements are the same in DR3 as in eDR3.

In addition, to avoid the selection of systems with bad astrometry or photometry, we limited the uncertainties in parallax and in observed fluxes using:
\begin{eqnarray}
    \varpi/\sigma_{\varpi} &>& 10 \\
    F_{BP}/\sigma_{F_{BP}} &>& 9 \\
    F_{RP}/\sigma_{F_{RP}} &>& 9
\end{eqnarray}
where $\sigma_{\varpi}$ is the parallax uncertainty, and $F_{BP/RP}$ and $\sigma_{F_{BP/RP}}$ are the reported fluxes and flux uncertainties in the $BP$/$RP$ passbands. The limits were tailored to recover AR~Sco, which despite its brightness shows relatively high $F_{BP}$ uncertainty due to variability ($F_{BP}/\sigma_{F_{BP}} = 9.943569$).

We also applied similar quality control criteria as in \citet{Pelisoli2019}, which were designed to exclude systems with bad parallax without removing known close binaries, namely:
\begin{eqnarray}
    E &<& 1.45 + 0.06\, (G_{BP} - G_{RP})^2 \\
    E &>& 1.0 + 0.015\,(G_{BP} - G_{RP})^2 \\
    u &<& 1.2\, \verb!max!(1, \exp(-0.2(G - 19.5)))
\end{eqnarray}
where $E$ is the excess noise when the $G$ band is compared to $G_{BP} + G_{RP}$ ($\verb!phot_bp_rp_excess_noise!$), and $u = \sqrt{ \verb!astrometric_chi2_al!/(\verb!astrometric_n_good_obs_al! - 5)}$, with
$\verb!astrometric_chi2_al!$ and $\verb!astrometric_n_good_obs_al!$ being, respectively, the value of the chi-square statistic of the astrometric solution and the number of good observations. The expression {\tt max} indicates that $u$ is compared with 1.2 times the largest value between 1 or $\exp(-0.2(G - 19.5))$.

This initial selection resulted in 28,124,276 objects, as illustrated in Fig.~\ref{fig:gaia}. Next we selected objects lying between the main sequence and the white dwarf cooling track, like AR~Sco. We applied the following selection:
\begin{eqnarray}
    M_G^{\prime} &<& 5.25 + 7\,((G_{BP} - G_{RP}) + 0.45)^{2/5} \\
    M_G^{\prime} &>&  4.45 - 5.15\,(1-\exp(0.35\,(G_{BP} - G_{RP})))
\end{eqnarray}
which is also shown in Fig.~\ref{fig:gaia}. Here we applied the median zero-point parallax of -0.017~mas to calculate $M_G^{\prime}$ \citep{Lindegren2021}. These cuts were tailored to approximately fit the 0.99 percentiles for the main sequence and for white dwarfs, thus excluding the vast majority of these systems from the selection. This {\it Gaia} selection resulted in 181,171 objects.

% Example figure
%\begin{figure}
	% To include a figure from a file named example.*
	% Allowable file formats are eps or ps if compiling using latex
	% or pdf, png, jpg if compiling using pdflatex
%	\includegraphics[width=\columnwidth]{example}
%    \caption{This is an example figure. Captions appear below each figure.
%	Give enough detail for the reader to understand what they're looking at,
%	but leave detailed discussion to the main body of the text.}
%    \label{fig:example_figure}
%\end{figure}

% Example table
%\begin{table}
%	\centering
%	\caption{This is an example table. Captions appear above each table.
%	Remember to define the quantities, symbols and units used.}
%	\label{tab:example_table}
%	\begin{tabular}{lccr} % four columns, alignment for each
%		\hline
%		A & B & C & D\\
%		\hline
%		1 & 2 & 3 & 4\\
%		2 & 4 & 6 & 8\\
%		3 & 5 & 7 & 9\\
%		\hline
%	\end{tabular}
%\end{table}

\subsection{WISE}

\begin{figure}
    \includegraphics[width=\columnwidth]{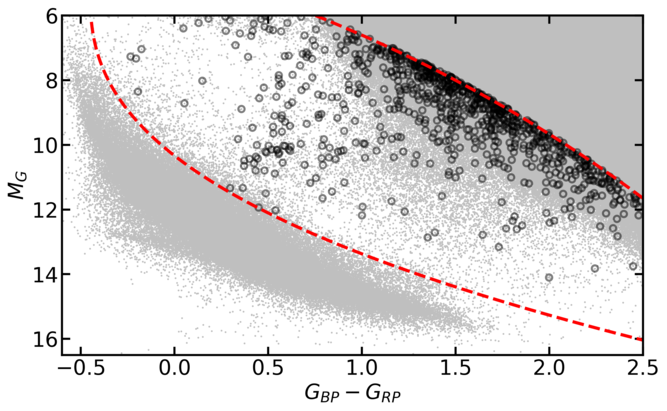}
    \caption{{\it Gaia} colour-magnitude diagram showing as black circles the 844 sources within our selected ranges (red dashed lines) flagged as variable. The grey points are the sources passing the initial colour-magnitude and quality selection (Eqs. 1-8).}
    \label{fig:wise_cmd}
\end{figure}

Another remarkable characteristic of AR~Sco is its infrared variability. Most white dwarfs are too faint in the infrared to be detected as variables, and therefore infrared variability can be a good initial identifier for binary white dwarf pulsars. To identify variable candidates, we matched the {\it Gaia} selection using a 5~arcsec radius with data from both the original (cryogenic) Wide-field Infrared Survey Explorer mission \citep[WISE,][]{wise} and from AllWISE \citep{allwise}, which combines the cryogenic and post-cryogenic \citep[NEOWISE,][]{neowise} survey phases. Both these catalogues contain variable flags that are 4-digit strings, with each digit corresponding to one of the WISE bands ($W_1$, $W_2$, $W_3$ and $W_4$, which are centred at 3.4~$\mu$m, 4.6~$\mu$m, 12~$\mu$m and 22~$\mu$m, respectively). The value of each digit goes from 0 to 9 (with $n$ for insufficient or inadequate data); the higher the number, the larger the probability of variability in a given band. Values of 5 or lower can generally be regarded as non-variable sources, 6--7 are potentially low amplitude variables, and 8--9 are likely variables. For AR~Sco, the flags are {\tt 9999} for WISE and {\tt 8866} for AllWISE. We selected as possible variables all objects whose flags contained a digit equal to or higher than 7. The angular resolution of WISE is, however, relatively poor ($6-12$~arcsec), making contamination by nearby sources common, which can lead to erroneous variability identification. Contamination is also flagged in both WISE and AllWISE with a four-digit string, similarly to the variability. A zero indicates no contamination, and different letters are used to indicate possible sources of contamination (for AR~Sco, the flags are {\tt 0000} in both WISE and AllWISE). We flagged as possibly contaminated all objects whose contamination flag contained {\tt h/H},  {\tt p/P}, {\tt d/D}, or  {\tt o/O} in either WISE or AllWISE. These indicate, respectively, potential contamination or spurious detection due to halo, persistent image, diffraction spikes, or optical ghosts caused by nearby stars. In this way, we identified 731 uncontaminated potential variable sources in WISE and 185 in AllWISE (with 102 common to both). We also flagged as variable objects those classified as variable in \citet{Petrosky2021} which had no contaminated flag (213 sources, 30 not flagged as variable in WISE or AllWISE). We therefore identified 844 candidate variables among the systems selected from {\it Gaia}, shown in Fig.~\ref{fig:wise_cmd}.

\begin{figure}
    \includegraphics[width=\columnwidth]{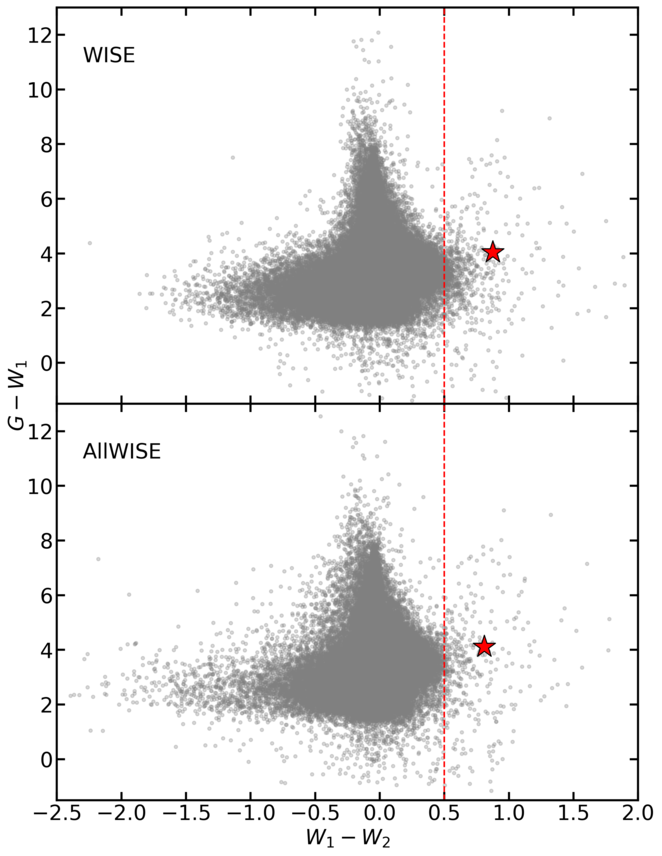}
    \caption{Colour-colour diagrams showing {\it Gaia} $G$ - $W_1$ versus $W_1 - W_2$ for objects in the {\it Gaia} selection that have WISE (top) or AllWISE (bottom) data. The red star shows the location of AR~Sco, which motivated the selection $W_1 - W_2 > 0.5$ (red dashed line).}
    \label{fig:wise_colour}
\end{figure}

Finally, AR~Sco's emission contains a significant contribution from non-thermal sources, unlike most systems lying on the same region of the {\it Gaia} colour-magnitude diagram, which we expect to be primarily detached white dwarf-main sequence binaries. We identified the WISE colour $W_1 - W_2$ as a good proxy for this, as AR~Sco clearly stands out from the bulk of other systems selected from {\it Gaia}, as shown in Fig.~\ref{fig:wise_colour}. Taking this into account, we applied a colour cut requiring either the WISE or the AllWISE value $W_1 - W_2$ to be larger than 0.5 to the variable sources, which resulted in 56 candidates, as well as recovering AR~Sco itself. All of the steps in the candidate selection are summarised in Table~\ref{tab:selection}.

\begin{table*}
	\centering
	\caption{Summary of the stages in our candidate selection.}
	\label{tab:selection}
	\begin{tabular}{lc} 
		\hline
		Total number of sources in Gaia eDR3 & 1,811,709,771 \\
		Sources after quality and initial colour-magnitude cuts (eqs. 1-8) & 28,124,276 \\
        Objects within the final colour-magnitude seleciton (eqs. 9-10) & 181,171 \\
        Systems showing WISE variability & 844 \\
        Candiates with $W_1 - W_2 > 0.5$ & 56 \\
		\hline
	\end{tabular}
\end{table*}

\section{Literature and Archival Data}

The 56 candidate binary white dwarf pulsars are listed in Table~\ref{tab:candidates}. 30 of these systems are known and previously well characterised as other types of sources in the literature. These are briefly discussed in Section~\ref{sec:known}. The other systems were not known or not well characterised, that is, they are only mentioned as candidates or detections. We first assessed the available archival data for these systems, as described in Section~\ref{sec:new}.

\begin{table*}
	\centering
	\caption{The 56 binary white dwarf pulsar candidates in our selection. Columns are the {\it Gaia} DR3 source identification, the Simbad ID (if the source has been previously well characterised), right ascension (RA) and declination (Dec) in J2000, and the {\it Gaia} apparent DR3 $G$ magnitude. We also include the GALEX $NUV$ magnitude when available, and the X-ray observatory that reported a detection. These two columns were not used in the candidate selection, but defined follow-up priority. Magnitudes are truncated reflecting precision. We also define a short name for each object which is used throughout the paper for simplicity. The last column indicates the classification of each system, preceded by $^{*}$ if it was determined as part of this work, as detailed in the text.}
	\label{tab:candidates}
	\begin{tabular}{ccccccccccc}
		\hline
		{\tt source\_id} & Simbad primary ID & RA & Dec & $G$ & $NUV$ & X-ray & Short name & Type\\
		\hline
%6050296829033196032 & V* AR Sco & 16:21:47.28 & -22:53:10.4 & 14.990 &  & XMM-NEWTON\\
4702994514879977472 &  & 00:07:43.23 & -69:59:47.7 & 19.153 & 17.726 & ROSAT & J0007-6959 & $^*$polar\\
4697621824327141248 & V* BL Hyi & 01:41:00.40 & -67:53:27.5 & 17.2134 & 17.657 & XMM-NEWTON & J0141-6753 & polar\\
4617143036371460864 &  & 01:56:37.23 & -83:58:34.2 & 16.697 & 17.260 & ROSAT & J0156-8358 & $^*$CV\\
5093945085525624448 & SSS~J035055.8-204817 & 03:50:55.99 & -20:48:15.8 & 14.775 & 17.627 & & J0350-2048 & YSO\\
5107845936158224768 &  & 03:54:10.31 & -16:52:50.0 & 17.346 & 18.048 & ROSAT & J0354-1652 & $^*$CV\\
3302990624137350528 & CRTS J035758.7+102943 & 03:57:58.66 & +10:29:42.4 & 18.102 &  & XMM-NEWTON & J0357+1029 & polar\\
473695898746498560 &  & 04:08:21.86 & +60:46:49.1 & 15.553 & 19.64 & & J0408+6046 & $^*$YSO\\
4871949697851411456 &  & 04:28:11.54 & -33:00:01.4 & 18.182 & 18.89 & & J0428-3300 & $^*$polar\\
4628229751415342976 &  & 04:35:09.62 & -75:27:44.0 & 16.912 & 17.424 & ROSAT & J0435-7527 & $^*$polar\\
159891122647000192 &  & 04:52:54.56 & +30:17:19.0 & 18.524 & 21.32 & & J0452+3017 & $^*$IP\\
4877265084954805504 & PPMXL 2431202680279124324 & 05:03:49.25 & -28:23:08.3 & 17.731 & 17.876 & XMM-NEWTON & J0503-2823 & IP\\
3223542525253775104 & 2MASS J05301240+0148214 & 05:30:12.40 & +01:48:21.4 & 16.612 &  & & J0530+0148 & YSO\\
4798833587650467200 & V* UW Pic & 05:31:35.65 & -46:24:05.0 & 15.861 & 18.099 & XMM-NEWTON & J0531-4624 & polar\\
3223900068396361856 & 2MASS J05315396+0242310 & 05:31:53.96 & +02:42:31.2 & 17.742 &  & & J0531+0242 & YSO\\
3016473390180542720 & 2MASS J05371640-0711463 & 05:37:16.41 & -07:11:46.4 & 17.481 &  &  & J0537-0711 & YSO\\
3334578455035280128 &  & 05:39:03.52 & +08:24:24.3 & 18.563 &  &  & J0539+0824 & $^*$YSO\\
3216423702860032896 & 2MASS J05405519-0247497 & 05:40:55.19 & -02:47:49.8 & 17.466 &  &  & J0540-0247 & YSO\\
3336335474617325568 & Gaia DR2 3336335474617325568 & 05:41:07.12 & +09:18:43.4 & 17.353 &  &  & J0541+0918 & YSO\\
3216159442110928256 & 2MASS J05422602-0308566 & 05:42:26.03 & -03:08:56.8 & 17.077 &  &  & J0542-0308 & YSO\\
3335057979544055168 &  & 05:51:28.99 & +07:52:41.0 & 17.808 &  &  & J0551+0752 & $^*$YSO\\
3100846613967311232 &  & 06:57:15.12 & -06:15:42.6 & 16.745 &  &  & J0657-0615 & $^*$YSO\\
3116059834803771904 & PPMXL 3189652360374206258 & 07:06:48.92 & +03:24:47.3 & 17.134 &  & XMM-NEWTON & J0706+0324 & polar\\
1101817445394854400 & V* HS Cam & 07:19:14.52 & +65:57:44.3 & 18.784 & 18.48 & XMM-NEWTON & J0719+6557 & polar\\
5532757148220186880 &  & 07:59:32.31 & -44:12:25.2 & 18.123 &  &  & J0759-4412 & $^*$YSO\\
3038572263932589696 &  & 08:08:06.87 & -09:35:13.6 & 18.4041 &  &  & J0808-0935 & unclear\\
5719598950133755392 & V* VV Pup & 08:15:06.80 & -19:03:17.8 & 15.955 &  & XMM-NEWTON & J0815-1903 & polar\\
5329108769930689664 & Gaia DR2 5329108769930689664 & 08:42:45.77 & -48:11:42.0 & 17.873 &  &  & J0842-4811 & YSO\\
790222752096510592 &  & 11:25:44.34 & +50:12:18.0 & 18.7939 & 18.743 &  & J1125+5012 & unclear\\
5788603818854455680 & [FLG2003] eps Cha 11 & 12:01:43.47 & -78:35:47.2 & 17.111 &  &  & J1201-7835 & YSO\\
3513017956589117056 &  & 12:26:37.90 & -23:04:14.5 & 19.536 & 22.97 &  & J1226-2304 & unclear\\
6086028036361965952 &  & 13:01:56.68 & -47:02:05.4 & 15.315 &  &  & J1301-4702 & $^*$YSO\\
6067533838480626432 &  & 13:05:47.97 & -55:02:34.0 & 16.249 &  &  & J1305-5502 & $^*$YSO\\
3718095055765579136 & BPS CS 30311-0012 & 13:25:22.38 & +06:00:28.8 & 14.538 & 18.75 &  & J1325+0600 & YSO\\
6096905573613586944 & V* V834 Cen & 14:09:07.29 & -45:17:16.0 & 16.609 &  & XMM-NEWTON & J1409-4517 & polar\\
5846719394999359232 &  & 14:25:40.45 & -68:37:32.9 & 18.406 &  &  & J1425-6837 & $^*$CV\\
5893562752219619200 & 2MASS J14534105-5521387 & 14:53:41.06 & -55:21:38.7 & 15.214 &  & INTEGRAL & J1453-5521 & polar\\
6243210920137332096 & CRTS J155929.1-223618 & 15:59:29.21 & -22:36:17.5 & 16.937 &  &  & J1559-2236 & YSO\\
6243223873758073728 & EPIC 204388640 & 16:02:04.30 & -22:31:46.9 & 17.209 &  &  & J1602-2231 & YSO\\
6242434969863194880 & 2MASS J16063093-2258150 & 16:06:30.92 & -22:58:15.1 & 18.475 &  &  & J1606-2258 & YSO\\
6048855674228736128 & Gaia DR2 6048855674228736128 & 16:17:44.78 & -25:09:15.5 & 17.882 &  &  & J1617-2509 & YSO\\
6033663795136741888 &  & 16:57:52.61 & -26:31:52.6 & 17.6845 &  &  & J1657-2631 & unclear\\
4476137370261520000 & V* V2301 Oph & 18:00:35.53 & +08:10:13.9 & 16.605 & 17.540 & XMM-NEWTON & J1800+0810 & polar\\
4585783762161656832 &  & 18:28:29.47 & +28:23:46.0 & 15.489 & 18.25 &  & J1828+2823 & $^*$YSO\\
4273036059117893632 & 2MASS J18294021+0015127 & 18:29:40.22 & +00:15:13.0 & 18.384 &  &  & J1829+0015 & YSO\\
4079708070610098432 &  & 18:45:53.17 & -21:46:34.9 & 17.644 &  &  & J1845-2146 & unclear\\
2262055164796424576 & V* EP Dra & 19:07:06.19 & +69:08:43.9 & 18.046 & 17.969 & XMM-NEWTON & J1907+6908 & polar\\
6712706405280342784 &  & 19:12:13.72 & -44:10:45.1 & 17.094 & 18.411 &  & J1912-4410 & pulsar\\
4186053835982000512 &  & 19:17:16.07 & -12:31:16.7 & 15.894 & 19.38 &  & J1917-1231 & $^*$YSO\\
6640165468504197120 & 2MASS J19293303-5603434 & 19:29:33.04 & -56:03:43.0 & 17.328 &  &  & J1929-5603 & polar\\
6909026160627539840 &  & 20:59:52.90 & -07:47:13.9 & 19.1572 &  &  & J2059-0747 & unclear\\
6883251168530612352 &  & 21:01:24.67 & -16:16:07.5 & 18.7265 &  &  & J2101-1616 & unclear\\
6827850007422236928 &  & 21:19:12.62 & -23:08:09.1 & 18.6381 &  &  & J2119-2308 & unclear\\
2222486906706076544 & 2MASS J21205785+6848183 & 21:20:57.85 & +68:48:18.4 & 17.018 &  &  & J2120+6848 & YSO\\
1956566510538468224 & Gaia DR2 1956566510538468224 & 22:04:50.69 & +40:08:38.4 & 18.498 & 19.79 & ROSAT & J2204+4008 & polar\\
1777232423131443840 & RX J2218.5+1925 & 22:18:32.76 & +19:25:20.5 & 17.312 & 18.053 & XMM-NEWTON & J2218+1925 & polar\\
1925212909978565632 &  & 23:36:11.32 & +44:25:39.6 & 16.907 &  &  & J2336+4425 & $^*$YSO\\
        \hline
	\end{tabular}
\end{table*}

\subsection{Known systems}
\label{sec:known}

Out of the 56 candidates identified in our initial search, 13 are known polars that have previously been well-characterised in the literature. All but one have orbital periods shorter than 3~hours. One system is an intermediate polar with spin and orbital periods of 23 and 81~min, respectively. Finally, 16 objects are characterised as young stellar objects (YSOs), such as T~Tauri stars. This large contamination likely arose due to our use of infrared variability as a criterion, which is also a characteristic of YSOs still surrounded by the remnants of the star-forming cloud. Notes and references on these 30 previously characterised systems are given in Appendix~\ref{apx:known} and their classes are given in Table~\ref{tab:candidates}. 

\subsection{Analysis of available archival data}
\label{sec:new}

The remaining 26 objects were not previously characterised in the literature, or had existing inconclusive characterisation. We first analysed archival photometric data that could potentially reveal pulses as those shown by AR~Sco. Specifically we analysed data from the Transiting Exoplanet Survey Satellite \citep[TESS,][]{tess}, the Zwicky Transient Facility \citep[ZTF,][]{ztf}, the Asteroid Terrestrial-impact Last Alert System \citep[ATLAS][]{atlas}, and the Catalina Real-Time Transient Survey \citep[CRTS,][]{crts}. The availability of data and the conclusions drawn from each dataset are summarised in Table~\ref{tab:photometry}

\begin{table*}
	\centering
	\caption{Summary of the analysis of archival photometric data for the 26 uncharacterised systems. Columns are the short name defined in Table~\ref{tab:candidates}, and the drawn conclusion (periodic, variable with no significant period, or not observed to vary -- NOV) based on TESS, ZTF, ATLAS, and CRTS is listed. For periodic systems, we also give the period from the dominant peak in hours. When the periodicity was found in more than one dataset, we report the period from TESS which was more precise in all cases. Periods from different surveys were always consistent.}
	\label{tab:photometry}
	\begin{tabular}{cccccc}
		\hline
		Short name & TESS & ZTF & ATLAS & CRTS & Period (h) \\
		\hline
  J0007-6959 & periodic & no data & variable & no data & $1.5826\pm0.0040$\\
  J0156-8358 & periodic & no data & variable & no data & $2.348\pm0.035$\\
  J0354-1652 & periodic & periodic & periodic & aperiodic & $1.690\pm0.034$\\
  J0408+6046 & NOV & variable & periodic? & no data & $1.71\pm0.14$\\
  J0428-3300 & periodic & no data & variable & NOV & $2.5466\pm0.0011$\\
  J0435-7527 & periodic & no data & periodic & variable & $2.36\pm0.23$\\
  J0452+3017 & periodic & periodic & periodic & no data & $1.325\pm0.008$\\
  J0539+0824 & NOV & variable & variable & variable & \\
  J0551+0752 & NOV & variable & variable & variable & \\
  J0657-0615 & variable & variable & variable & no data & \\
  J0759-4412 & NOV & no data & variable & no data & \\
  J0808-0935 & NOV & NOV & NOV & no data & \\
  J1125+5012 & NOV & variable & variable & no data & \\
  J1226-2304 & NOV & NOV & NOV & NOV & \\
  J1301-4702 & periodic? & no data & variable & variable & $54\pm7$\\
  J1305-5502 & periodic & no data & variable & no data & $35.2\pm0.9$\\
  J1425-6837 & NOV & no data & variable & no data & \\
  J1657-2631 & no data & NOV & NOV & no data & \\
  J1828+2823 & variable & variable & variable & no data & \\
  J1845-2146 & no data & NOV & NOV & no data & \\
  J1912-4410 & periodic & no data & periodic & variable & $4.037\pm0.006$\\
  J1917-1231 & no data & variable & variable & no data & \\
  J2059-0747 & no data & NOV & NOV & no data & \\
  J2101-1616 & no data & NOV & NOV & NOV & \\
  J2119-2308 & NOV & NOV & NOV & NOV & \\
  J2336+4425 & variable & variable & variable & variable & \\
        \hline
	\end{tabular}
\end{table*}

\subsection{TESS}

% summary of TESS data; how many objects have available data and how many were variable out of those.

The TESS mission monitored more than 85 per cent of the sky. Each sector -- a pre-defined 24$^{\circ}$ $\times$ 90$^{\circ}$ region of the sky -- is monitored for 27 days consecutively, with a brief interruption halfway through for data downlinking. There is overlap between different sectors, such that the total baseline can be as long as 351 continuous days (for stars around the ecliptic poles). The cadence of available light curves is either 20~s or 2~min, and 10~min or 30~min images are available for all observed sectors. The pixel size is 21'', meaning that contamination from nearby stars is a potential issue.

No 20-sec light curves were available for the 26 uncharacterised objects. Three out of the 26 objects (J0354-1652, J0408+6046 and J1828+2823) have available 2-min light curves. We used specifically the light curves provided by the TESS Science Processing Operations Center (SPOC) and the PDCSAP flux, which corrects an initial simple aperture photometry (SAP) to remove instrumental trends and contributions from neighbouring stars other than the target of interest using a pre-search data conditioning (PDC). All but five systems have 10/30-min images. We performed the photometry for these images using {\sc eleanor} \citep{eleanor} in a custom-built script \footnote{https://github.com/ipelisoli/eleanor-LS}.

\begin{figure*}
    \centering
    \begin{minipage}{.5\textwidth}
        \includegraphics[width=0.98\textwidth]{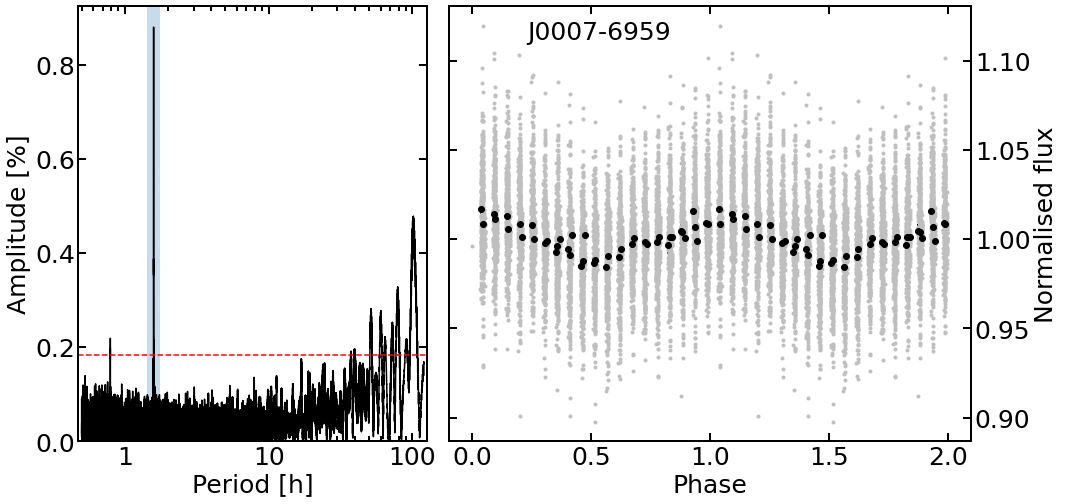}
        \includegraphics[width=0.98\textwidth]{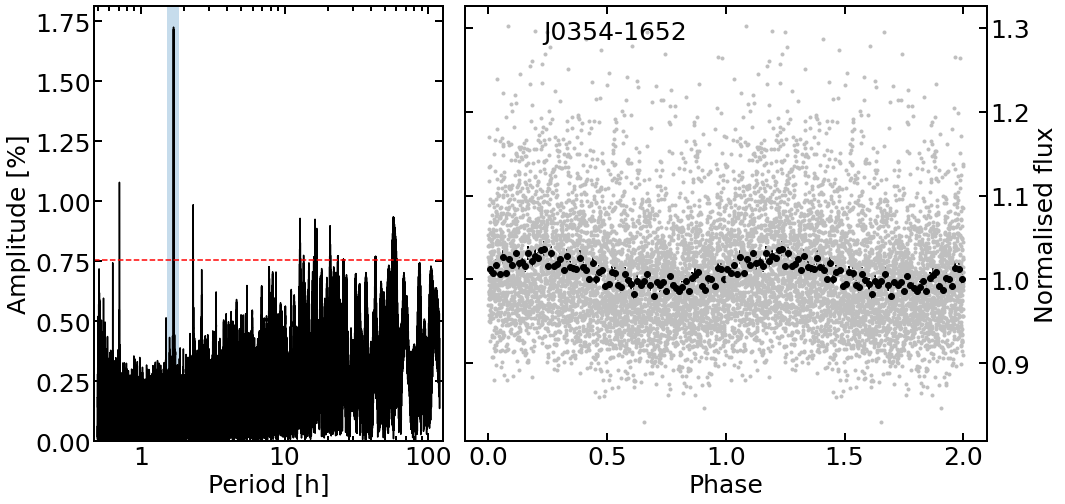}
        \includegraphics[width=0.98\textwidth]{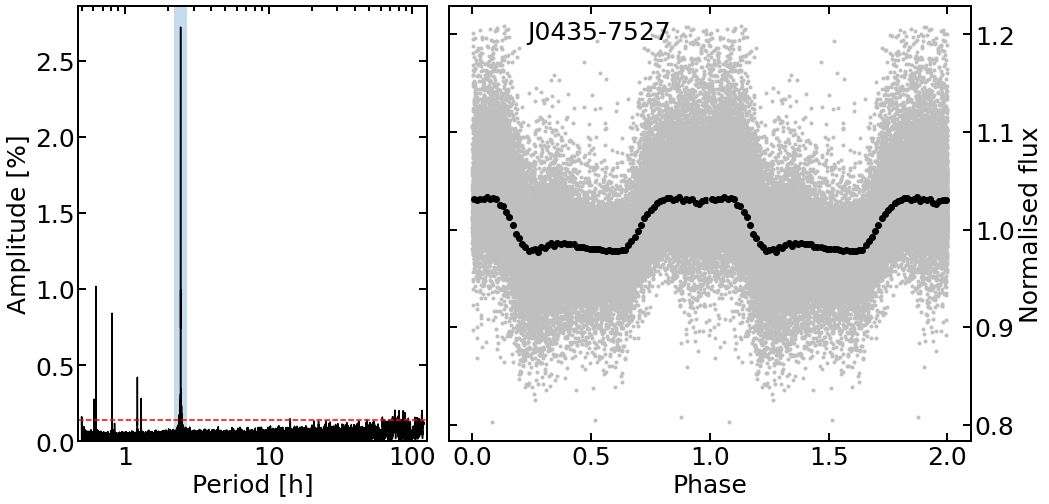}
        \includegraphics[width=0.98\textwidth]{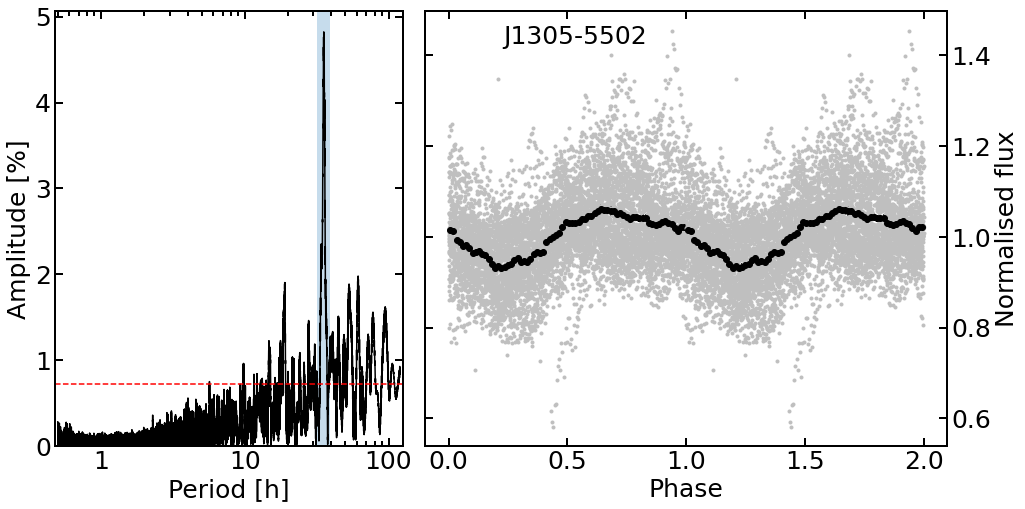}
    \end{minipage}%
    \begin{minipage}{0.5\textwidth}
        \centering
        \includegraphics[width=0.98\textwidth]{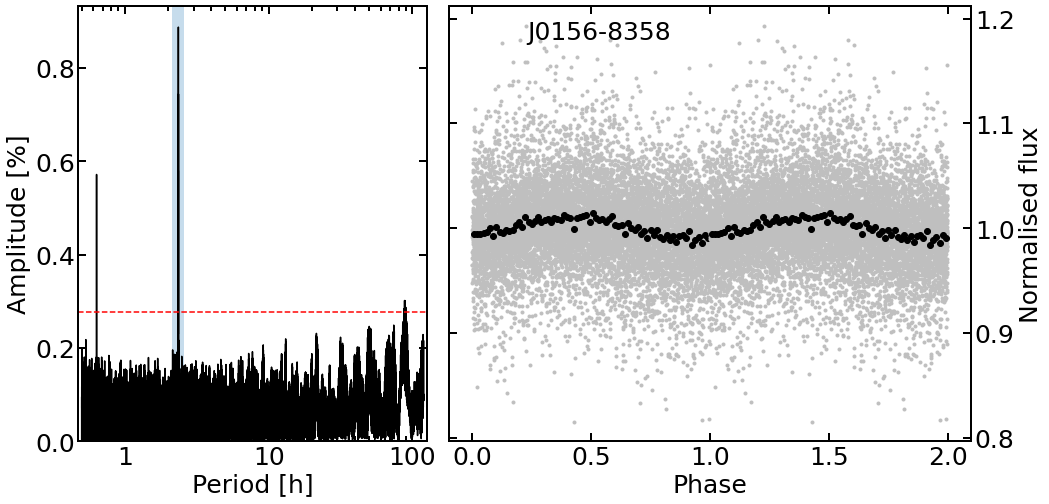}
        \includegraphics[width=0.98\textwidth]{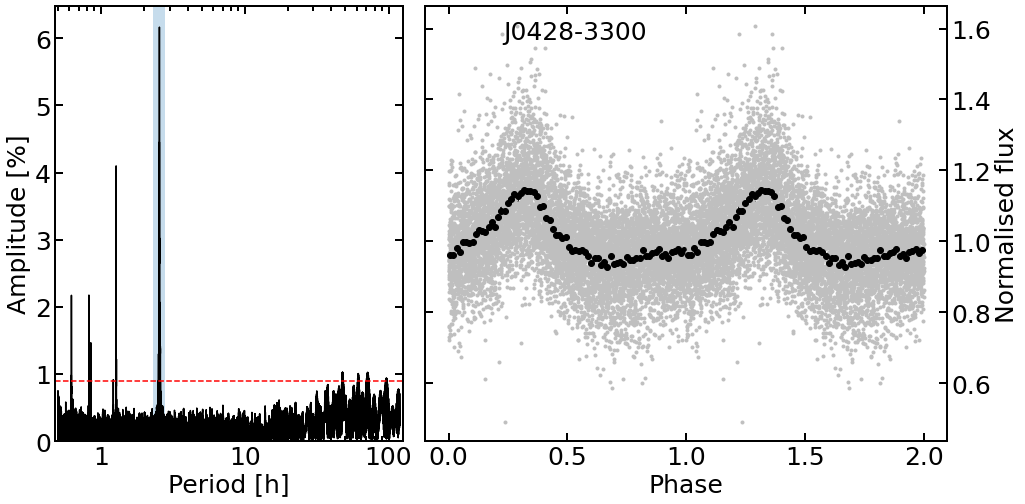}
        \includegraphics[width=0.98\textwidth]{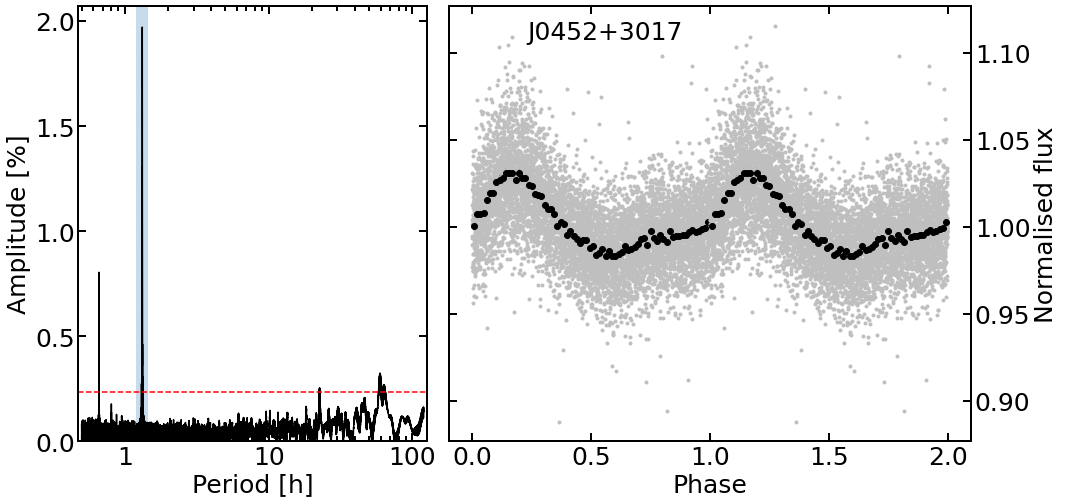}
        \includegraphics[width=0.98\textwidth]{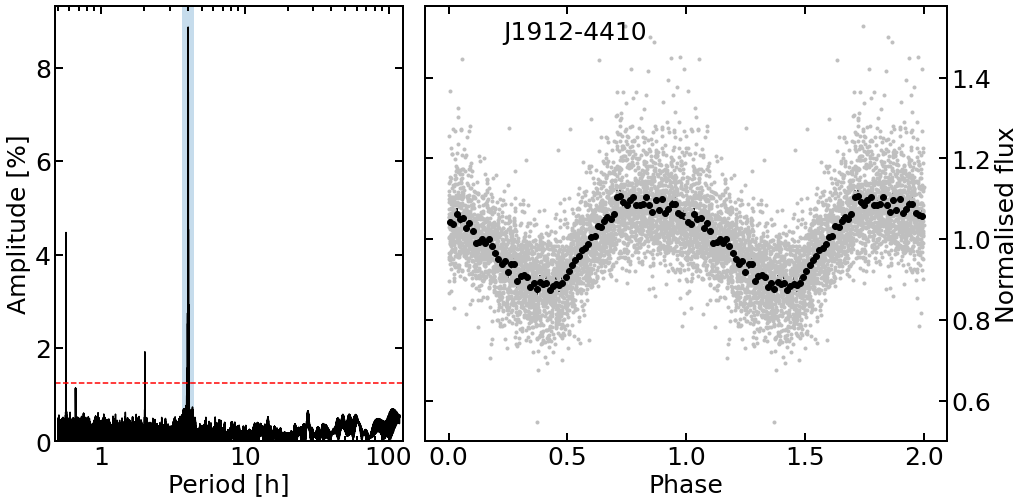}
    \end{minipage}
    \caption{The eight systems showing periodic variability in  TESS data. In each case, the left panel shows the periodogram, with the horizontal red dashed line indicating an average 5-$\sigma$ detection limit. The blue shaded area indicates the dominant period, with the data folded to that period showed on the right panels in grey dots. The black filled circles show the data averaged to 100 phase bins (the zeropoint of phase was arbitrary at this stage).}
    \label{fig:tess_periodic}
\end{figure*}

\begin{figure}
    \centering
    \includegraphics[width=\columnwidth]{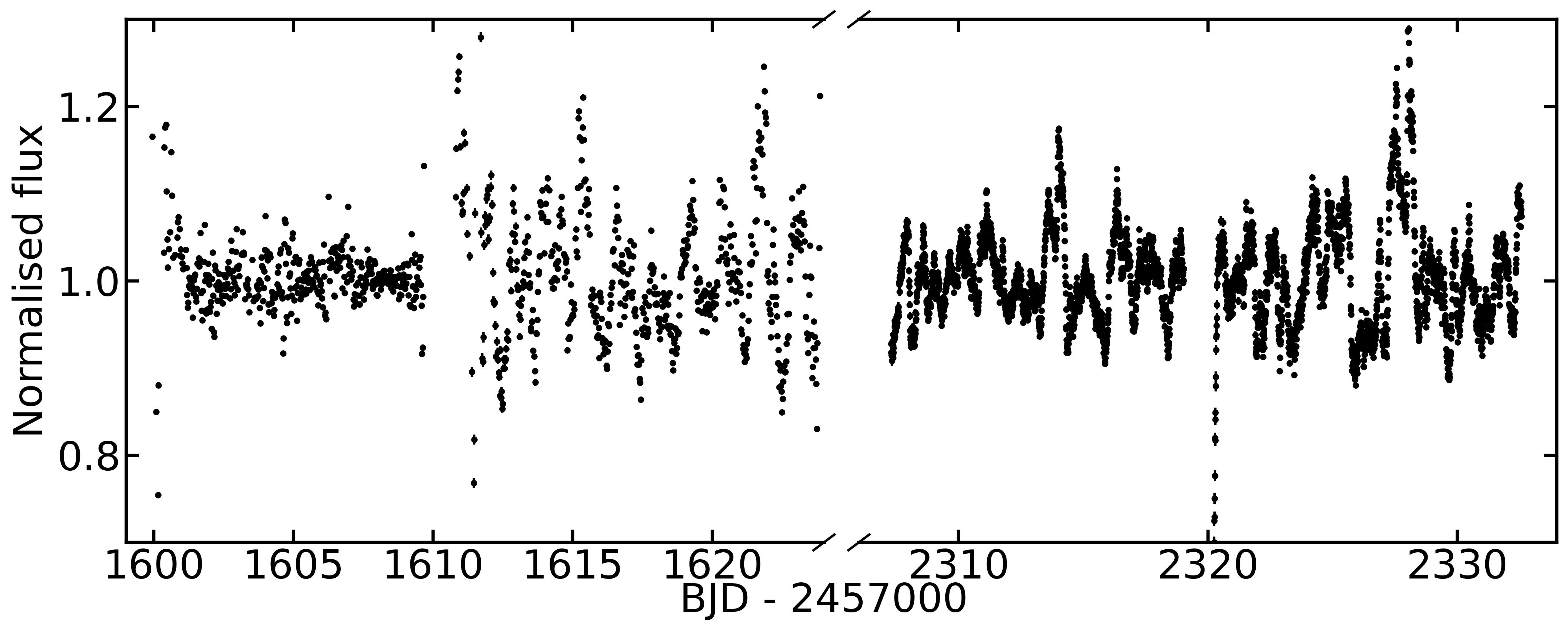}
    \includegraphics[width=0.46\textwidth]{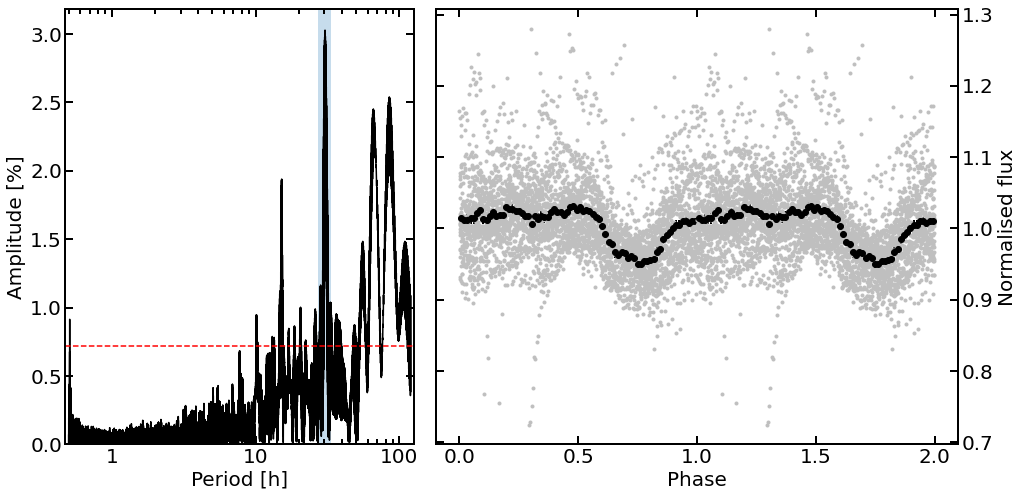}
    \caption{J1301-4702, which shows both aperiodic variability (clear in the top panel) and potential periodic variability as suggested in the bottom panels (left is the periodogram, right the phase folded light curve). The potential period (indicated by the blue shading) is above a 1 per cent false alarm probability calculated by randomising the flux measurements and recalculating the periodogram.}
    \label{fig:tess_both}
\end{figure}

Out of the 21 targets with available data, eight show clear periodic variability (see Fig.~\ref{fig:tess_periodic}), with an additional system potentially showing a periodicity on top of stochastic behaviour (Fig.~\ref{fig:tess_both}). The obtained periods range from 1.3 to over 50~h. Performing pre-whitening revealed no independent periods -- all other periodicities could be attributed to harmonics or cadence aliases. Given the periods and light curve shapes, the most probable causes for variability are binarity (irradiation or eclipses) or rotation; no pulsing behaviour like that of AR~Sco is seen in the light curves. Three systems (J0452+3017, J0428-3300, and J1912-4410) display saw-tooth-shape light curves with periods of the order of hours, similar to the orbital behaviour shown by AR~Sco. One of those is J1912-4410, which we have confirmed as a binary white dwarf pulsar from follow-up observations \citep{Pelisoli2023, Schwope2023}.

\begin{figure}
    \includegraphics[width=\columnwidth]{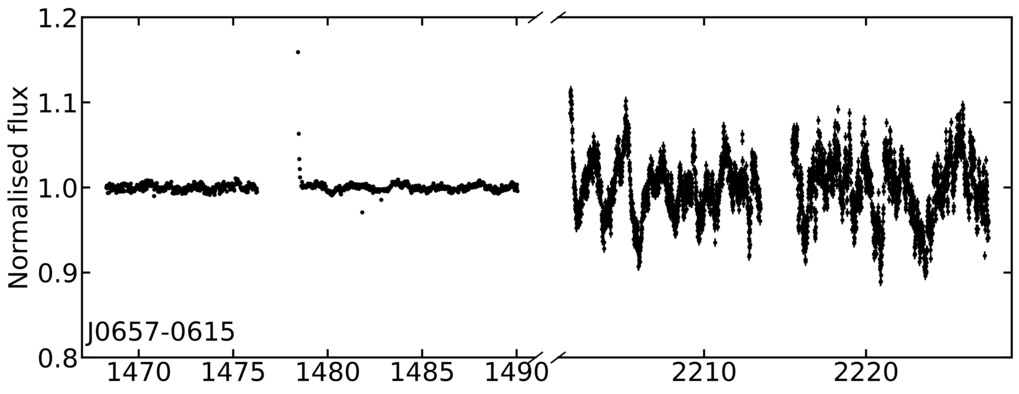}
    \includegraphics[width=\columnwidth]{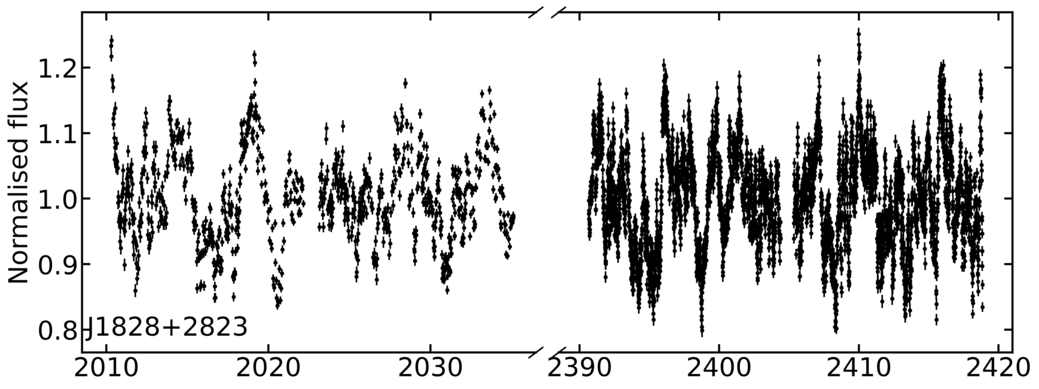}
    \includegraphics[width=\columnwidth]{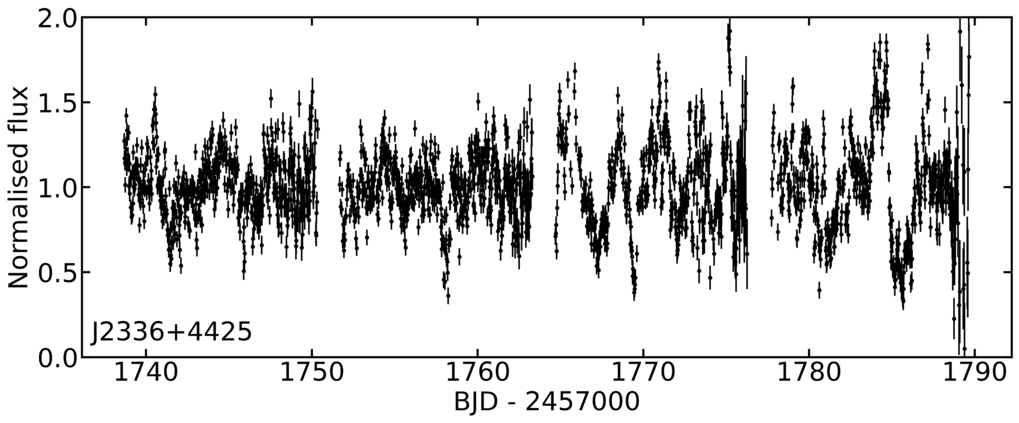}
    \caption{Light curves for the three systems showing aperiodic variability in TESS data. The big change in amplitude for J0657-0615 could be due to a change in state between the two observations; however, given the large pixel size of TESS, the change could also be explained by a change in the contribution from other sources to the aperture.}
    \label{fig:tess_aperiodic}
\end{figure}

Three other systems also show aperiodic variability, as shown in Fig.~\ref{fig:tess_aperiodic}. Our adopted criterion for aperiodic variability is the same used by \citet{Pelisoli2024b}, where the distribution of standard deviations from the mean is compared with the maximum expected deviation from the mean given the number of observations. Gaussian noise is assumed to calculate the maximum expected deviation, and a threshold of more than 10 per cent of the observations above the expected deviation (to allow for systematics) is used to define variability. The cause for aperiodic variability could be, for example, changes in accretion rate in systems that have a disc, such as cataclysmic variables (CVs) and YSOs. The other nine systems show no detectable periodic or aperiodic variability.

\subsection{ZTF}

% summary of ZTF data; how many objects have available data and how many were variable out of those.

ZTF uses the 1.2-m Palomar telescope to monitor the northern sky (declination $\gtrsim -25^{\circ}$). The exposure time is 30~sec and observations use primarily the $g$ and $r$ filters, with occasional observations in the $i$ band. Images are usually taken every three nights, with some fields observed for longer in "deep drilling" events. Photometry is made available through the NASA/IPAC Infrared Science Archive.

\begin{figure}
    \includegraphics[width=\columnwidth]{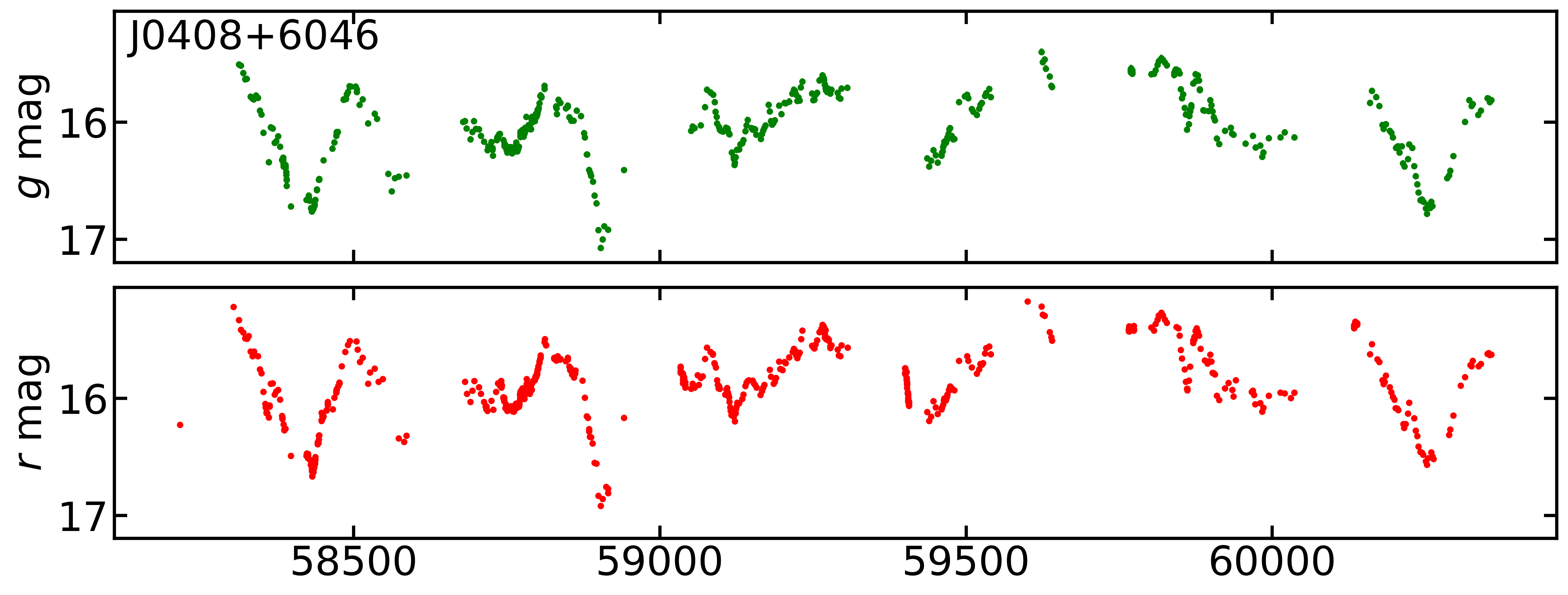}
    \includegraphics[width=\columnwidth]{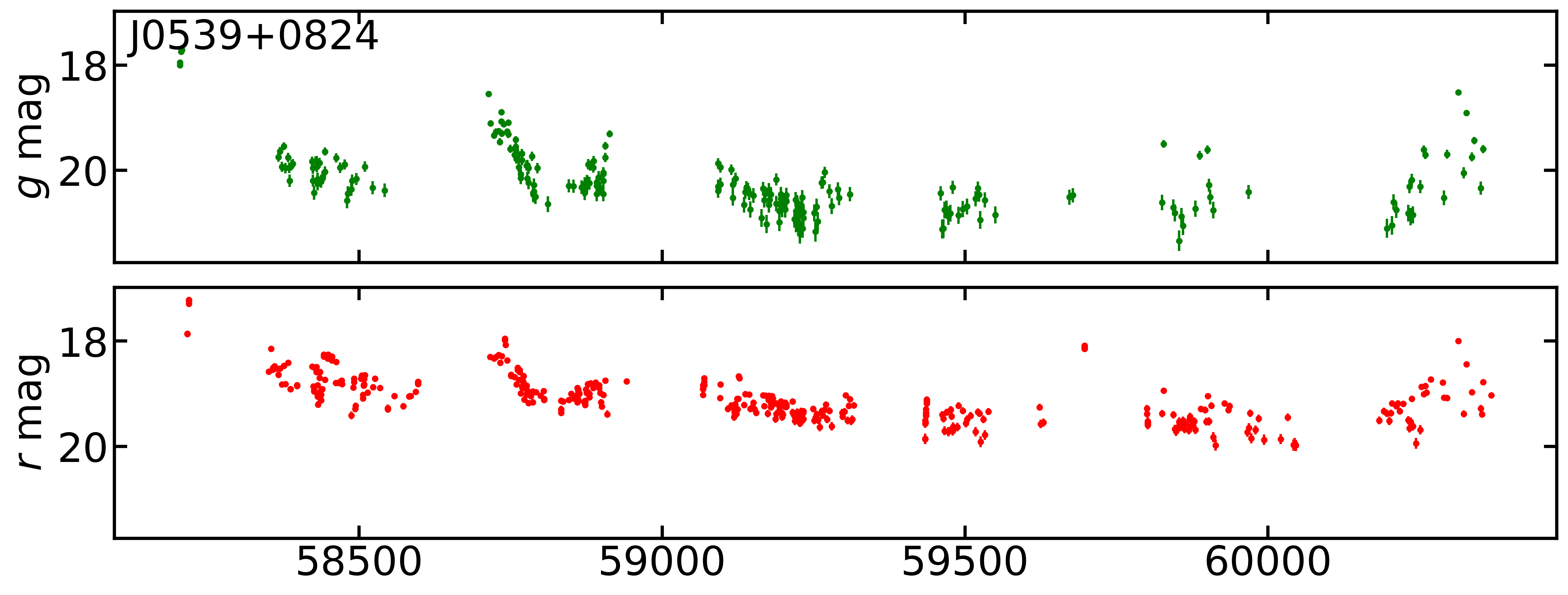}
    \includegraphics[width=\columnwidth]{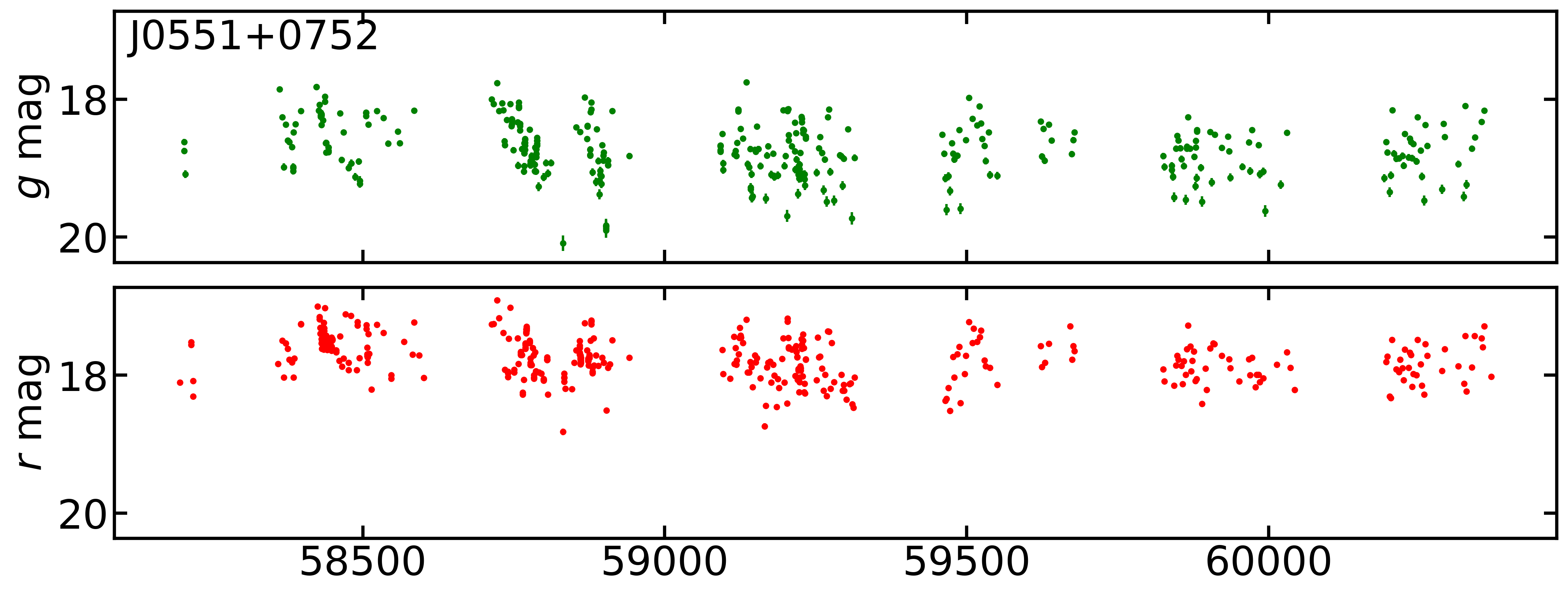}
    \includegraphics[width=\columnwidth]{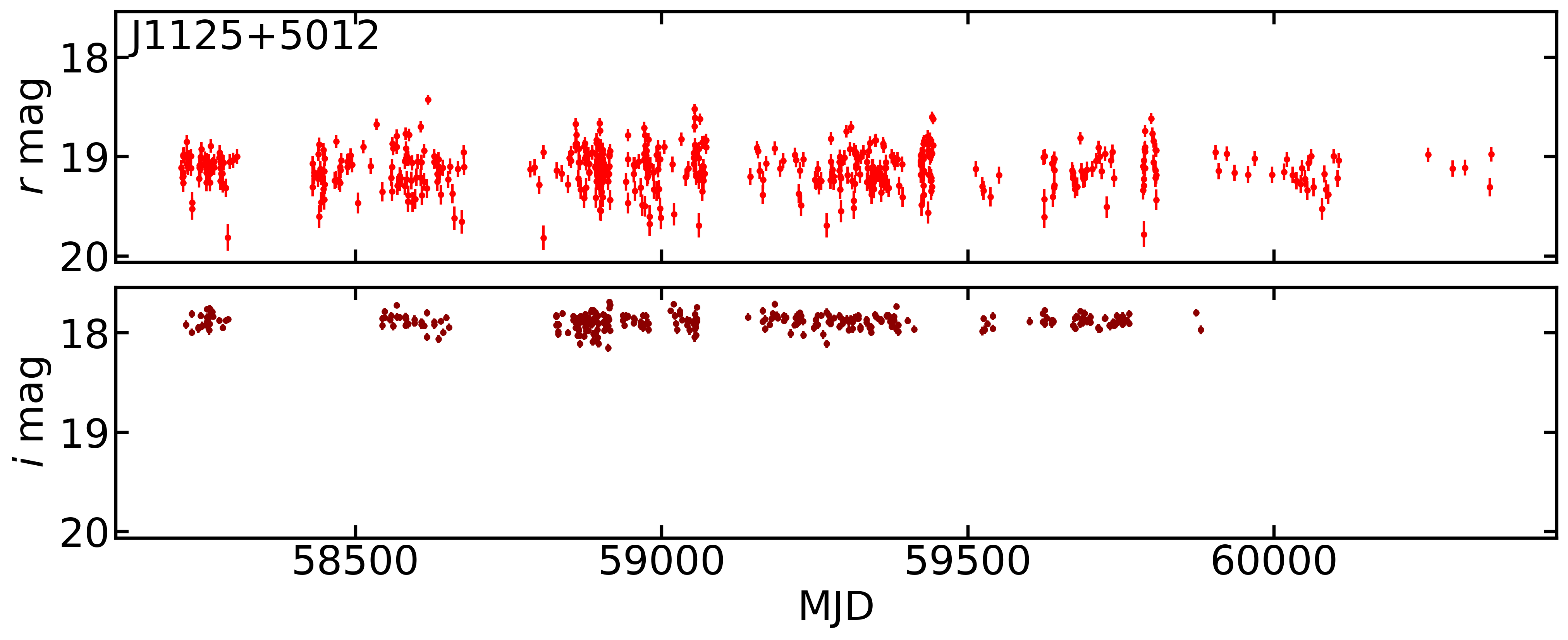}    \includegraphics[width=\columnwidth]{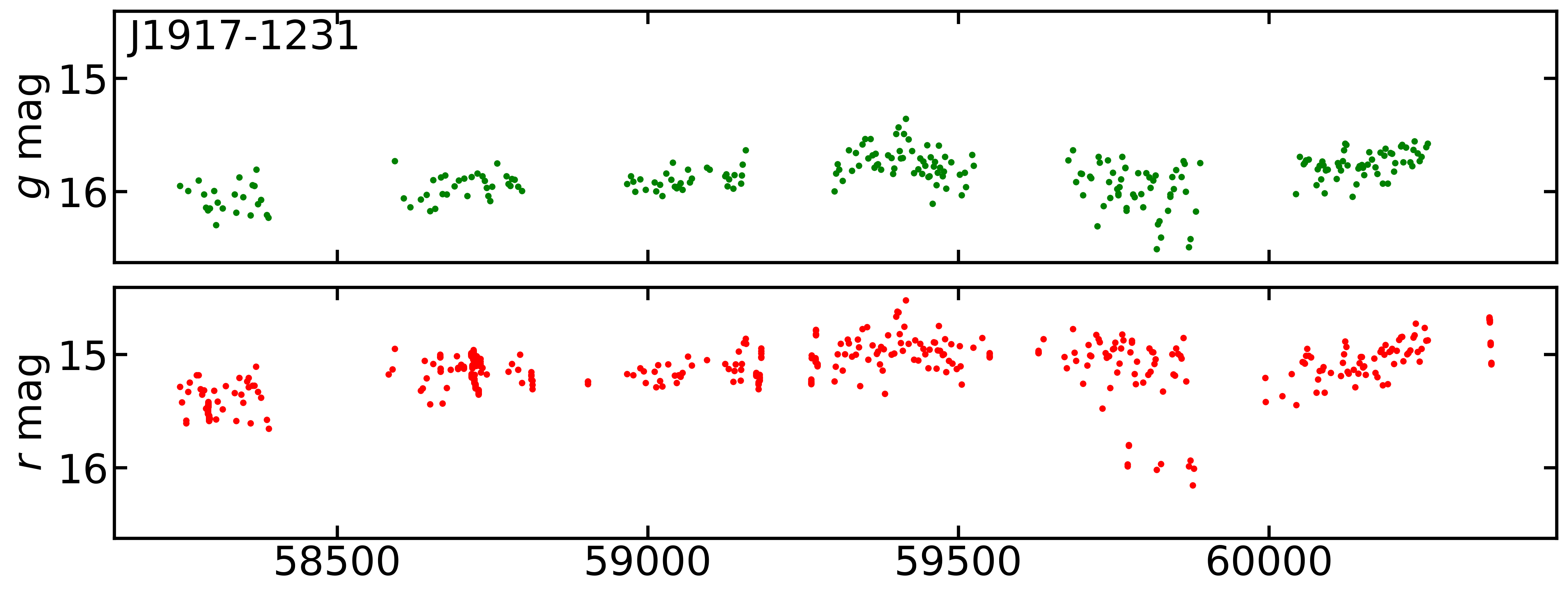}
    \caption{Light curves for the five systems that show aperiodic variability in ZTF.}
    \label{fig:ZTF_aperiodic}
\end{figure}

17 out of the 26 uncharacterised systems have ZTF data. All of these have $r$ band data and all but one (J1125+5012) have $g$ data. Six have $i$ band data, but only one of these (J1125+5012, the same system that has no $g$ data) has a substantial number of measurements ($>100$). To search for periodic signals, we calculated a multi-band periodogram as described by \citet{VanderPlas2015} using {\tt gatspy} \citep{gatspy}, which implements a weighted sum of individual Lomb-Scargle periodograms. We used the $g$ and $r$ data, except in the case of J1125+5012 where $r$ and $i$ were used. Two systems (J0354-1652 and J0452+3017) were found to show periodic behaviour with periods consistent with those found in the TESS data. Like previously, pre-whitening revealed no additional periodicities. It is worth noting that none of the other systems found to be periodic with TESS have ZTF data. Eight systems are aperiodic according to the metric described previously; four of these appeared to be constant in TESS -- possibly because of flux dilution from nearby stars, or due to longer-term variability than the TESS baseline -- and one had no TESS data. These five systems whose variability had not been revealed by TESS are shown in Fig.~\ref{fig:ZTF_aperiodic}. Seven systems showed no detectable periodic or aperiodic variability.

\subsection{ATLAS}

% summary of ATLAS data; how many objects have available data and how many were variable out of those.

ATLAS utilises four 0.5-m telescopes (two in Hawaii, one in Chile, one in South Africa) to scan the sky, primarly to detect Near Earth Objects. The exposures are 30-sec long and use broad orange ($o$) and cyan ($c$) filters. The whole sky is scanned down to a magnitude of about 19.5 with a cadence of one day for declinations between -50 and +50$^{\circ}$ or two days in the polar regions. To obtain photometric measurements for the 26 systems, we employed the forced photometry server \footnote{https://fallingstar-data.com/forcedphot/}.

\begin{figure}
    \includegraphics[width=\columnwidth]{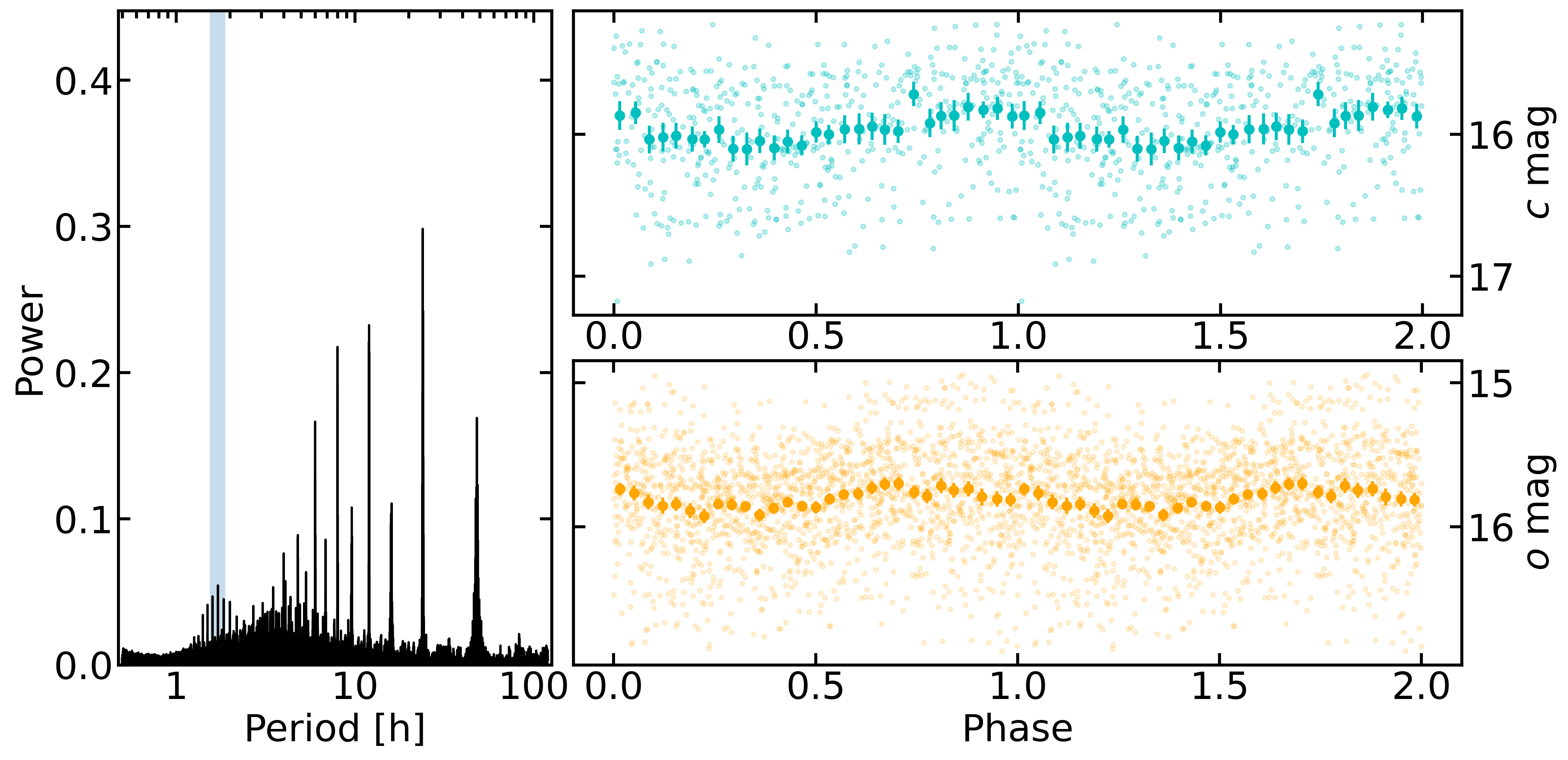}
    \caption{J0408+6046, the system showing periodic variability detected only with ATLAS. The left panel shows the periodogram which is dominated by one-day aliases; the possible period is marked by the shaded blue region. The $c$ (top) and $o$ (bottom) light curves folded to this period are shown on the right panels.}
    \label{fig:atlas_periodic}
\end{figure}

\begin{figure}
    \includegraphics[width=\columnwidth]{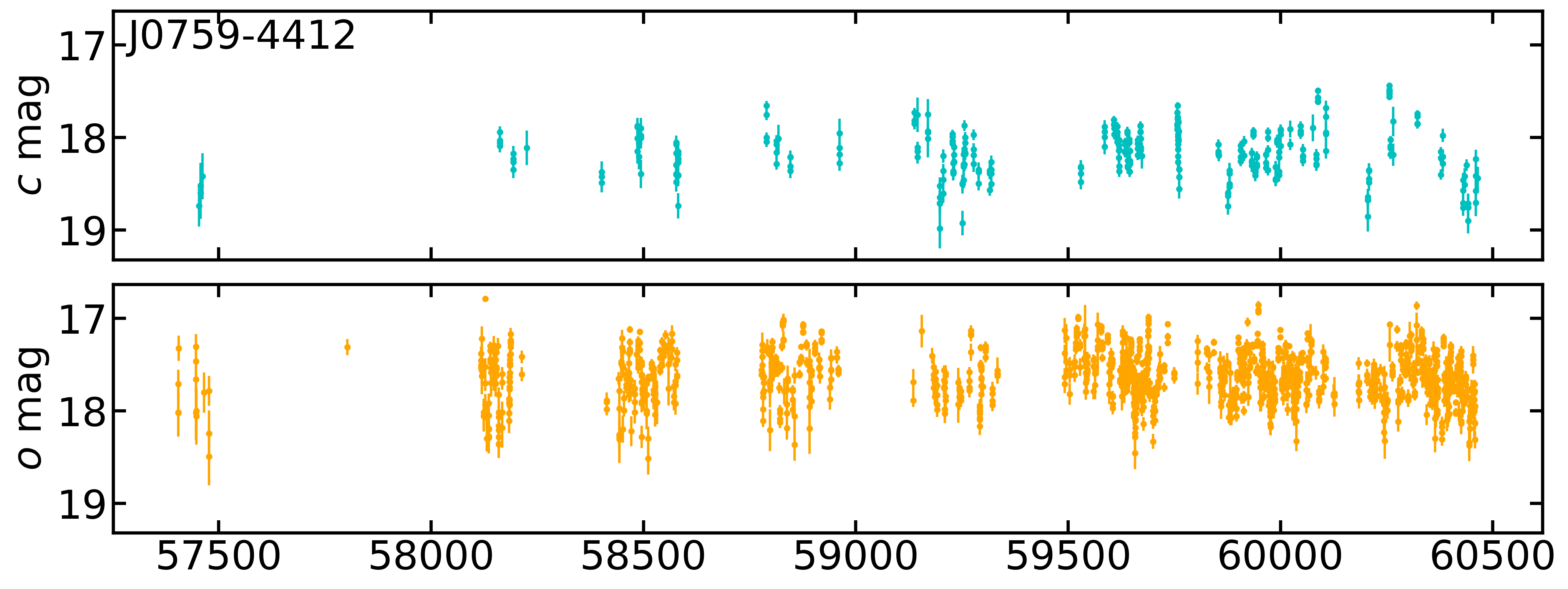}
    \includegraphics[width=\columnwidth]{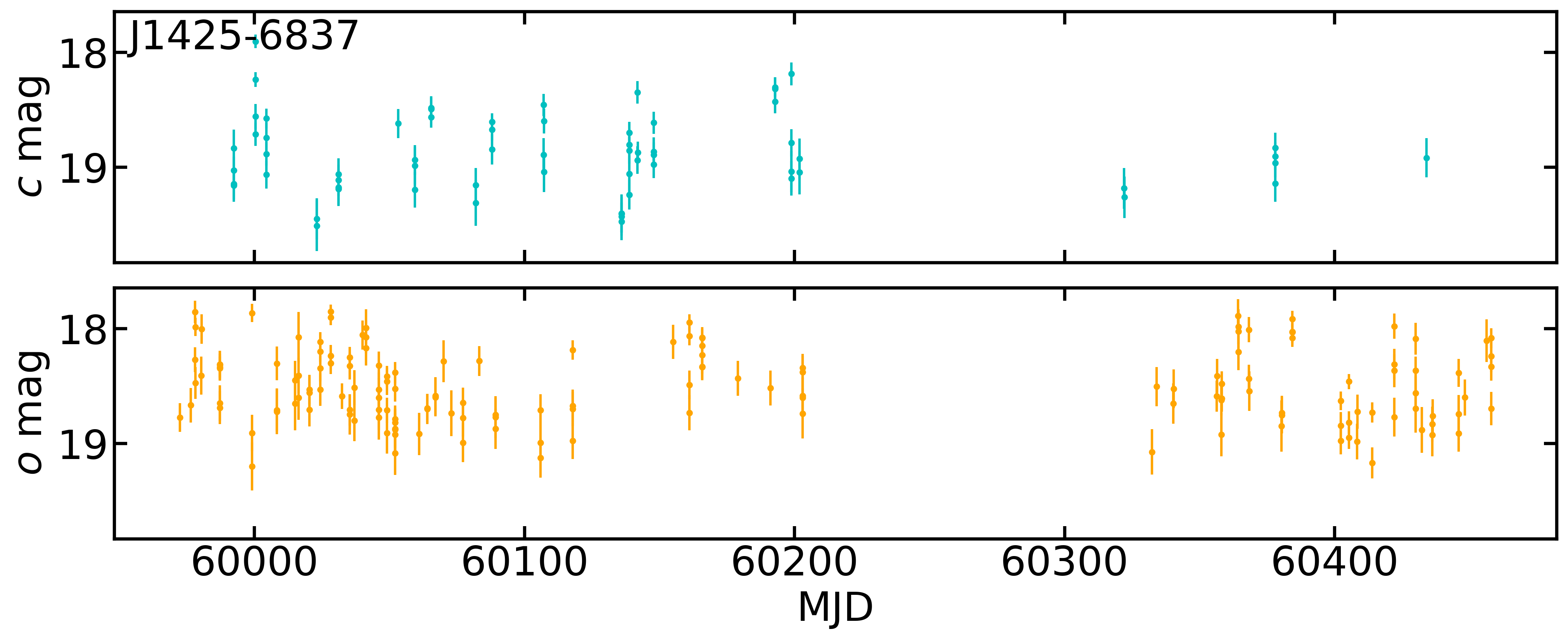}
    \caption{Light curves for the two systems showing aperiodic variability not detected in TESS but detected in ATLAS.}
    \label{fig:atlas_aperiodic}
\end{figure}

We analyse the data in the same way as the ZTF data, by doing a multi-band periodogram and by searching for aperiodic variability. We find five systems showing detectable periodic variability, four of which show periods consistent with those found by TESS and a fifth that was not detected as periodic by TESS or ZTF (though aperiodic variability was seen in ZTF). The light curve for this system is shown in Fig.~\ref{fig:atlas_periodic}. The periodogram is dominated by one-day aliases, but an independent periodicity can also be seen. Additionally, two systems found to be constant with TESS and that had no ZTF data appear to be variable in ATLAS. They are shown in Fig.~\ref{fig:atlas_aperiodic}. Results for other systems are consistent with the TESS and ZTF findings, that is periodic systems were found to be variable even if the period was undetermined, and systems appearing constant in TESS and ZTF were also found to be constant in ATLAS.

\subsection{CRTS}

% summary of CRTS data; how many objects have available data and how many were variable out of those.

CRTS uses three telescopes, two in the northern hemisphere (the 0.7-m Catalina Sky Survey telescope at Mt. Bigelow Station, and the 1.5-m Mt. Lemmon Survey telescope) and one in the southern hemisphere (the 0.5-m Siding Springs Survey telescope, which was discontinued in 2013). No filters are used, but magnitudes are made available in an approximate $V$ scale. Typically four 30-sec exposures are obtained per night, reaching a limiting magnitude around 19--21.

CRTS data are available for 11 out of the 26 uncharacterised objects. No new variables are revealed by this dataset. Seven systems met our criterion for aperiodic variability (all of which were also found to vary with other surveys), whereas the remaining four systems showed no detectable variability.

\section{Follow-up observations and analysis}

\subsection{High-speed time-series photometry}

% table with all ucam observations

The cadence and exposure time of photometric surveys is not ideal for identifying fast pulsed variability like that seen for AR~Sco on a 2-min timescale. Therefore, most of the 26 uncharacterised systems were also followed up with the high-speed photometers ULTRACAM \citep{ultracam} and ULTRASPEC \citep{ultraspec}. ULTRACAM is mounted on the 3.5-m European Southern Observatory (ESO) New Technology Telescope (NTT) and is equipped with a beam splitter that allows simultaneous observations in three filters. ULTRASPEC is installed on the 2.4-m Thai National Telescope (TNT) and observes only one band at a time. In both cases, the readout time is negligible ($\lesssim 0.5$~ms) thanks to frame-transfer capabilities. The journal of observations is detailed in Table~\ref{tab:fastphot}. All data were reduced using the HiPERCAM pipeline \footnote{https://cygnus.astro.warwick.ac.uk/phsaap/hipercam/docs/html/}. We carried out bias subtraction, flat-field correction, and then aperture photometry using a constant star as reference. The same star was used when observations were repeated in more than one night.

\begin{table*}
	\centering
	\caption{Journal of ULTRACAM and ULTRASPEC observations with the object's short name, the instrument used, filters, date the night started, the cadence and (in case of ULTRACAM) number of co-added $u$ images, and the duration of the observation. J1912-4410 is omitted given that detailed ULTRACAM follow-up is described in \citet{Pelisoli2023}.}
	\label{tab:fastphot}
	\begin{tabular}{ccccccc}
		\hline
		{\tt source\_id} & Instrument & Filter(s) & Date & Cadence (s) & $u$ co-adds & Duration (min) \\
		\hline
  J0007-6959 & ULTRACAM & $u$, $g$, $r$ & 2024-05-03 & 7.0 & 3 & 35\\
  \\
  J0156-8358 & ULTRACAM & $u$, $g$, $i$ & 2022-03-06 & 3.0 & 3 & 38\\
  \\
  J0354-1652 & ULTRASPEC & $g$ & 2021-02-14 & 4.8 & - & 150\\
                      & ULTRASPEC & $g$ & 2021-02-15 & 4.8 & - & 104\\
                      & ULTRACAM & $u$, $g$, $i$ & 2021-03-07 & 2.5 & 3 & 108\\
  \\                    
  J0428-3300 & ULTRACAM & $u$, $g$, $i$ & 2021-03-10 & 6.0 & 3 & 49\\
                      & ULTRACAM & $u$, $g$, $i$ & 2021-03-17 & 3.1 & 3 & 131\\
                      & ULTRACAM & $u$, $g$, $i$ & 2021-07-08 & 5.0 & 3 & 81\\
                      & ULTRACAM & $u$, $g$, $i$ & 2021-11-08 & 3.0 & 4 & 173\\
  \\
  J0435-7527 & ULTRACAM & $u$, $g$, $i$ & 2021-01-22 & 5.8 & 3 & 98\\
                      & ULTRACAM & $u$, $g$, $i$ & 2021-03-08 & 5.8 & 3 & 175\\
  \\
  J0452+3017  & ULTRASPEC & $g$ & 2023-02-23 & 9.9 & - & 184\\
  \\
  J0539+0824 & ULTRACAM & $u$, $g$, $r$ & 2023-03-18 & 8.0 & 3 & 61\\
  \\
  J0551+0752 & ULTRACAM & $u$, $g$, $i$ & 2023-10-11 & 5.8 & 3 & 41\\
  \\
  J0657-0615 & ULTRASPEC & $g$ & 2022-02-07 & 4.9 & - & 61\\
                      & ULTRACAM & $u$, $g$, $i$ & 2023-03-09 & 4.0 & 3 & 106\\
  \\
  J0759-4412 & ULTRACAM & $u$, $g$, $i$ & 2023-03-12 & 6.5 & 3 & 16\\
                      & ULTRACAM & $u$, $g$, $r$ & 2023-03-19 & 4.0 & 3 & 69\\
                      & ULTRACAM & $u$, $g$, $i$ & 2023-03-29 & 3.4 & 3 & 84\\
  \\
  J0808-0935 & ULTRACAM & $u$, $g$, $r$ & 2023-04-23 & 5.1 & 3 & 31\\
  \\
  J1226-2304 & ULTRACAM & $u$, $g$, $r$ & 2023-04-23 & 8.1 & 5 & 47\\
                      & ULTRACAM & $u$, $g$, $r$ & 2024-02-09 & 5.7 & 3 & 103\\
  \\
  J1301-4702 & ULTRACAM & $u$, $g$, $i$ & 2021-03-09 & 1.0 & 4 & 70\\
                      & ULTRACAM & $u$, $g$, $i$ & 2021-03-10 & 2.4 & 2 & 46\\
  \\
  J1305-5502 & ULTRACAM & $u$, $g$, $i$ & 2022-06-05 & 8.0 & 3 & 94\\
  \\
  J1425-6837 & ULTRACAM & $u$, $g$, $i$ & 2022-06-06 & 9.0 & 3 & 39\\
                      & ULTRACAM & $u$, $g$, $i$ & 2022-06-07 & 5.1 & 3 & 24\\
                      & ULTRACAM & $u$, $g$, $r$ & 2023-04-23 & 8.1 & 3 & 110\\
                      & ULTRACAM & $u$, $g$, $i$ & 2023-04-29 & 5.0 & 3 & 36\\
  \\
  J1657-2631 & ULTRACAM & $u$, $g$, $i$ & 2022-06-07 & 4.1 & 5 & 32\\
                      & ULTRASPEC & KG5 & 2023-04-22 & 8.2 & - & 52\\
  \\
  J1845-2146 & ULTRACAM & $u$, $g$, $r$ & 2023-03-08 & 4.0 & 3 & 47\\
                      & ULTRACAM & $u$, $g$, $i$ & 2023-03-20 & 4.0 & 4 & 62\\
  \\
  J1917-1231 & ULTRACAM & $u$, $g$, $i$ & 2022-06-05 & 3.4 & 6 & 36\\
                     & ULTRASPEC & KG5 & 2023-04-22 & 3.1 & - & 41\\
  \\
  J2059-0747 & ULTRACAM & $u$, $g$, $i$ & 2023-10-17 & 6.1 & 3 & 64\\
  \\
  J2101-1616 & ULTRACAM & $u$, $g$, $r$ & 2023-04-23 & 5.8 & 5 & 61\\
                      & ULTRACAM & $u$, $g$, $r$ & 2023-10-18 & 5.8 & 3 & 31\\
  \\
  J2119-2308 & ULTRACAM & $u$, $g$, $r$ & 2023-09-14 & 5.8 & 3 & 28\\
  \\
  J2336+4425 & ULTRASPEC & $g$ & 2022-12-18 & 5.6 & - & 110\\
        \hline
	\end{tabular}
\end{table*}
 
\begin{figure}
    \includegraphics[width=\columnwidth]{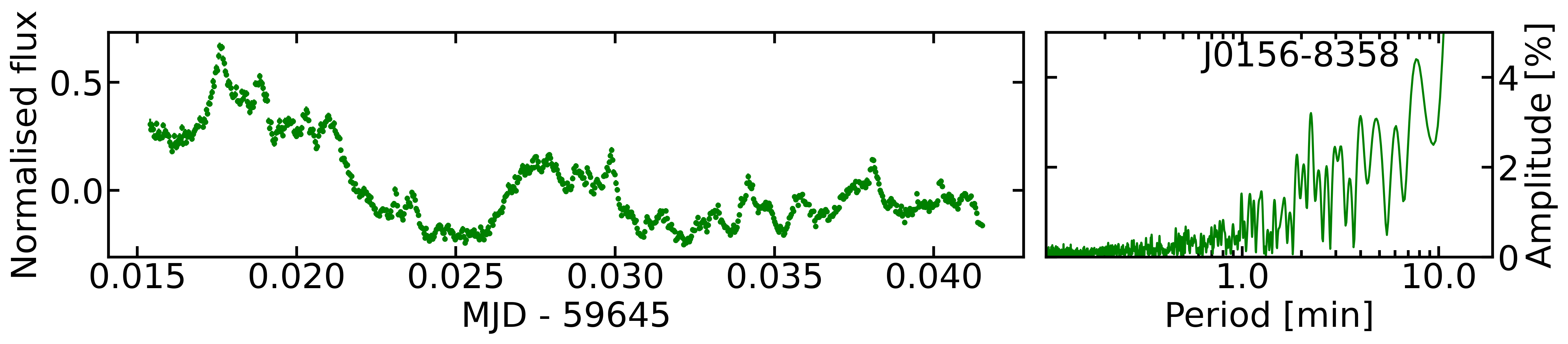}
    \includegraphics[width=\columnwidth]{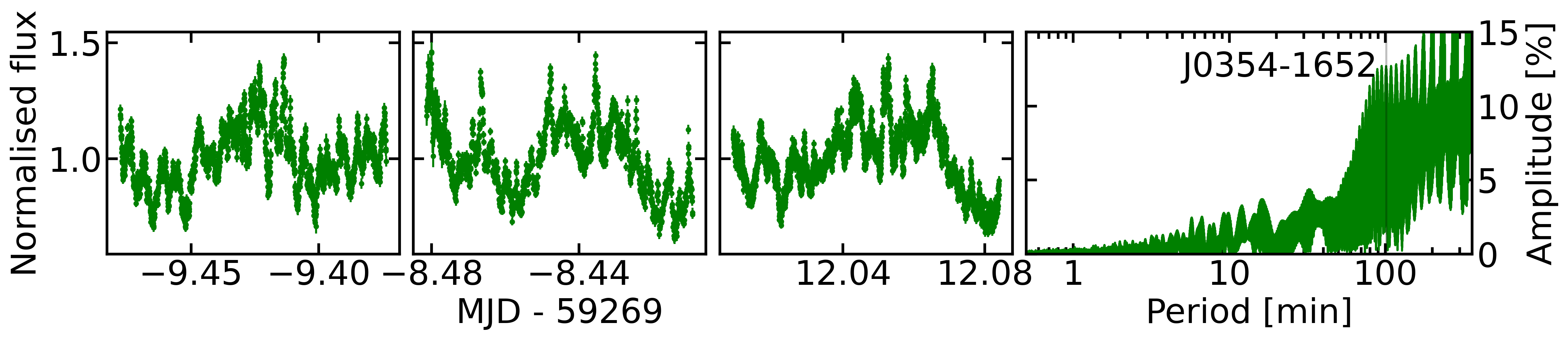}
    \caption{The $g$ band light curves (left panels) and periodogram (rightmost panel) for J0156-8358 (top) and J0354-1652 (bottom). No periods stand out from the background. The period detected in archival data for J0354-1652 is indicated by a vertical dashed line.}
    \label{fig:ucam_stochastic}
\end{figure}

\begin{figure*}
    \includegraphics[width=0.8\textwidth]{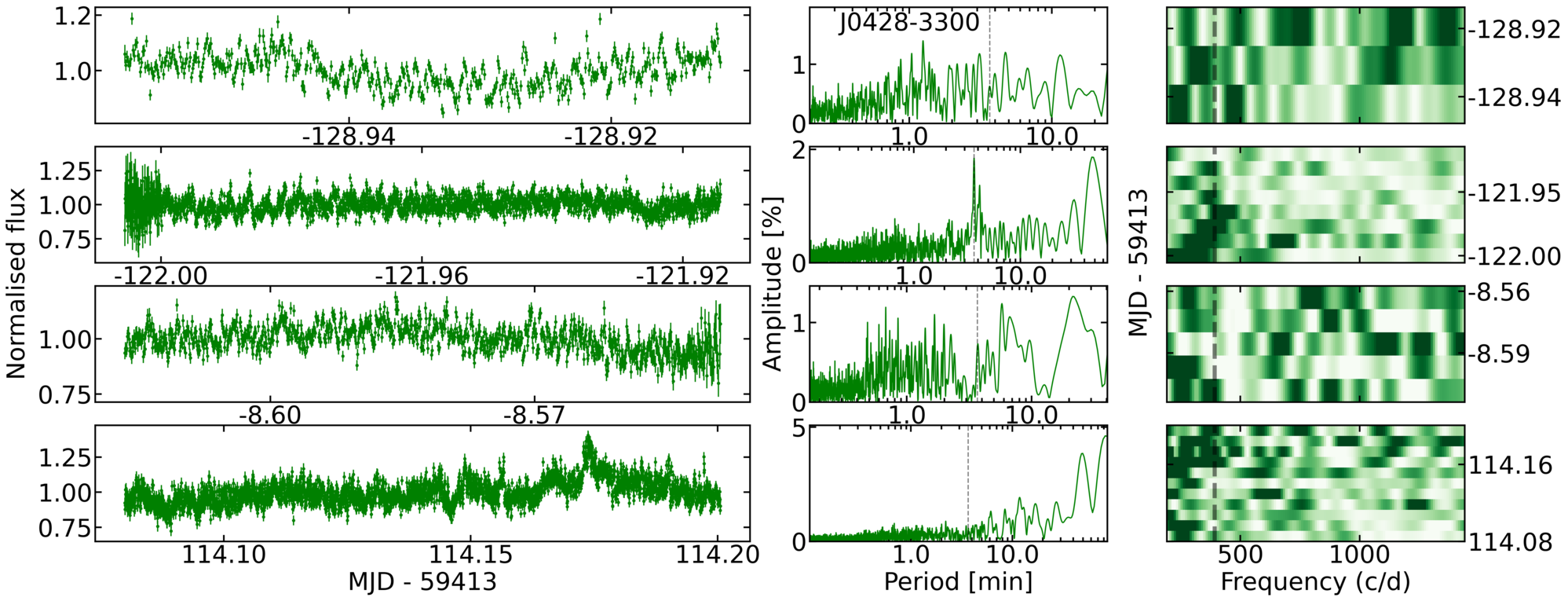}
    \includegraphics[width=0.8\textwidth]{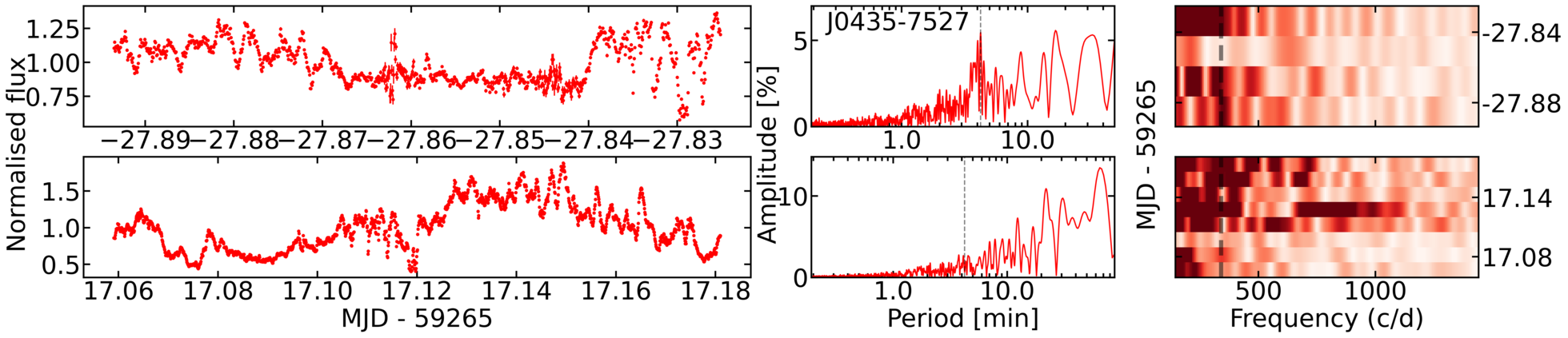}
    \caption{The light curves (left panels), periodogram (middle) and running Fourier transform (right) for the $g$ data of J0428-3300 (top four panels, in green) and $r$ data of J0435-7527 (bottom two panels, in red). Both systems show a peak in the periodogram (indicated by dashed vertical lines) that does not seem to be persistent, as it is not always detected and drifts in the running Fourier transform.}
    \label{fig:ucam_qpos}
\end{figure*}

\begin{figure}
    \includegraphics[width=\columnwidth]{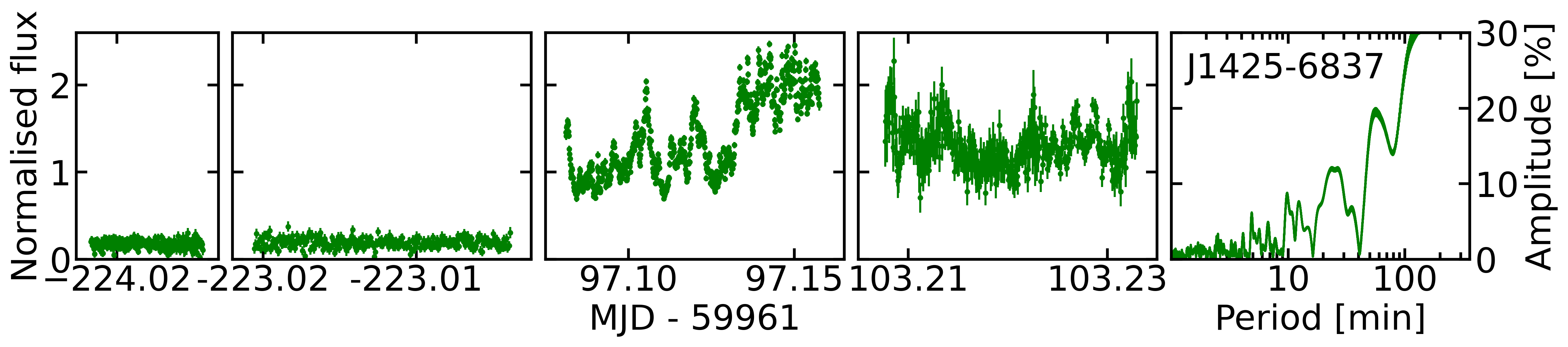}
    \caption{$g$-band light curves (left panels) and periodogram (rightmost panel, only including the two last nights of data) of J1425-6837. In the first two nights the system has very low signal and was barely above the readout noise. The following year it was over a magnitude brighter and displaying variability.}
    \label{fig:ucam_j1425}
\end{figure}

\begin{figure}
    \includegraphics[width=\columnwidth]{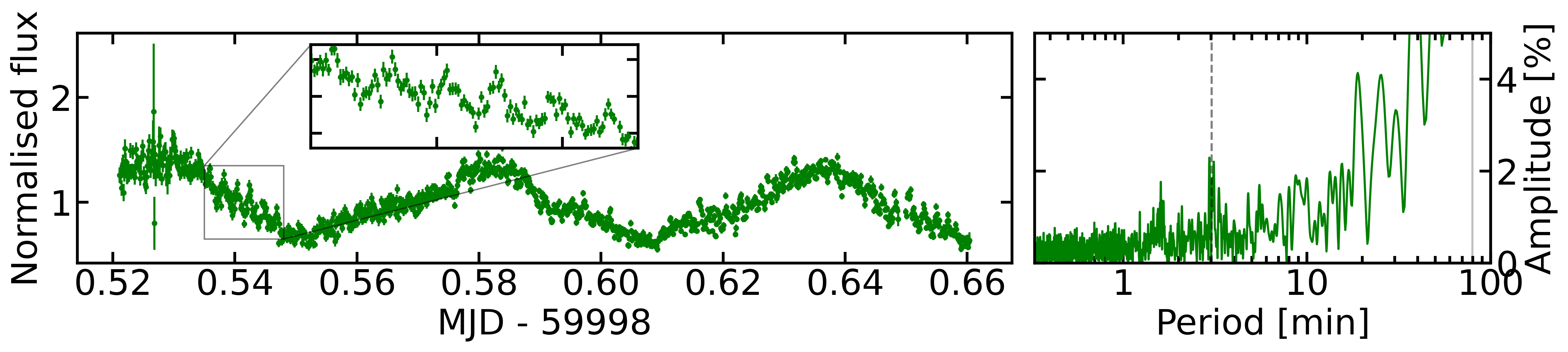}
    \caption{The $g$ light curve of J0452+3017 (left panel), with an inset zooming in on the visible pulses. The right panel shows the periodogram with the newly detected period indicated by a vertical dashed line, and the period found with archival data shown by the solid grey line.}
    \label{fig:ucam_j0452}
\end{figure}

Six systems showed fast and seemingly stochastic variability in their light curves, typical of CV systems. For two of these, J0156-8358 and J0354-1652, Fourier analysis revealed no periodic variability (see Fig.~\ref{fig:ucam_stochastic}). J0428-3300 and J0435-7527 show potential quasi-periodic oscillations, whose frequency drifts and sometimes is not even detected. For J0428-3300 that translates to a periodogram peak at $3.676\pm0.009$~min, and for J0435-7527 there is a peak $4.24\pm0.11$~min (see Figure~\ref{fig:ucam_qpos}). J1425-6837, was constant and barely detectable in two observations, and then detected and stochastically variable the next year, suggesting it might be a dwarf nova (see Figure~\ref{fig:ucam_j1425}) with the abrupt brightness change caused by a sudden dumping of matter onto the white dwarf due to instability in the disk \citep[e.g.][]{Hameury2020}. Finally, J0452+3017 showed a hint of a period at $3.0\pm0.5$~min with visible pulsing behaviour (see Fig.~\ref{fig:ucam_j0452}), akin to AR~Sco albeit with lower amplitude. This behaviour prompted dedicated time-resolved spectroscopy follow-up (Section~\ref{sec:GTC}).

\begin{figure}
    \includegraphics[width=\columnwidth]{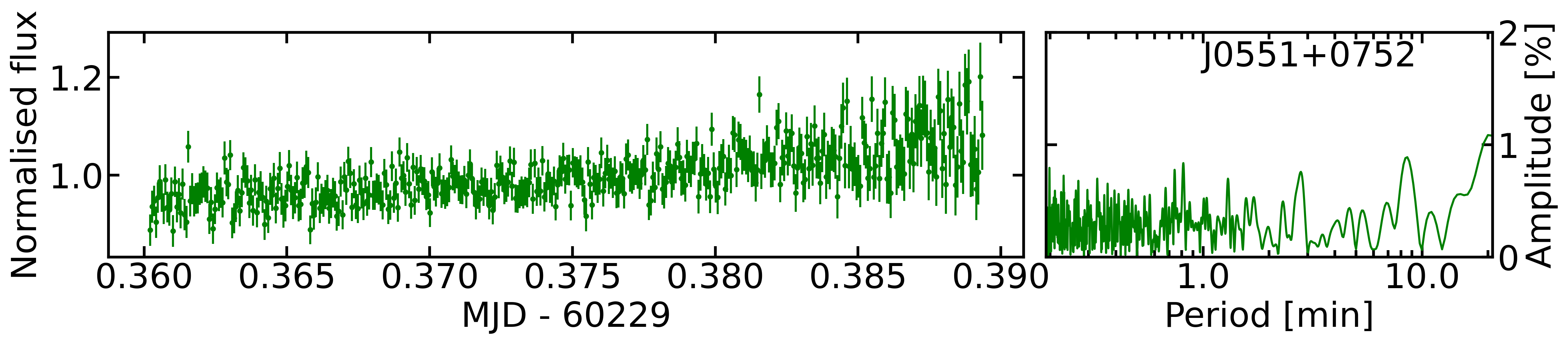}
    \includegraphics[width=\columnwidth]{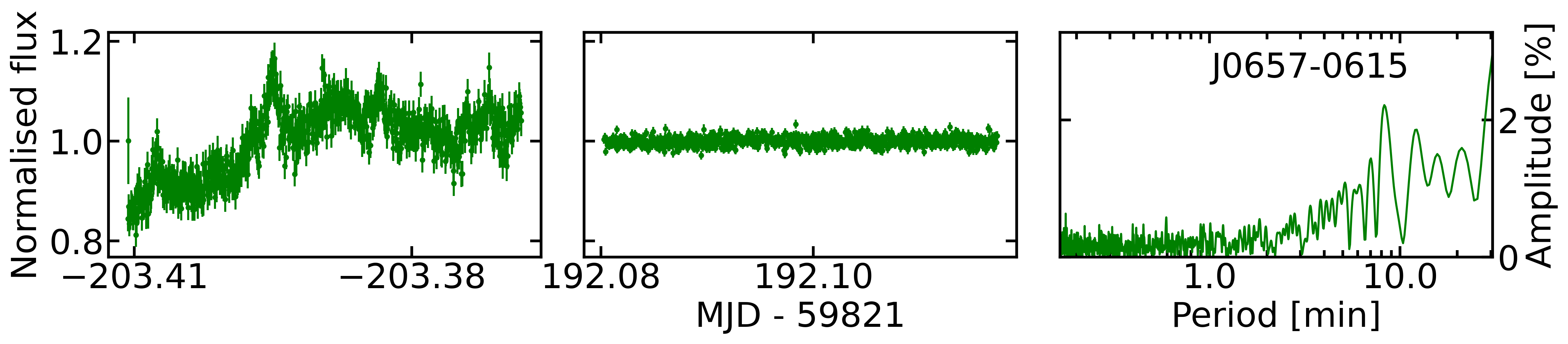}
    \includegraphics[width=\columnwidth]{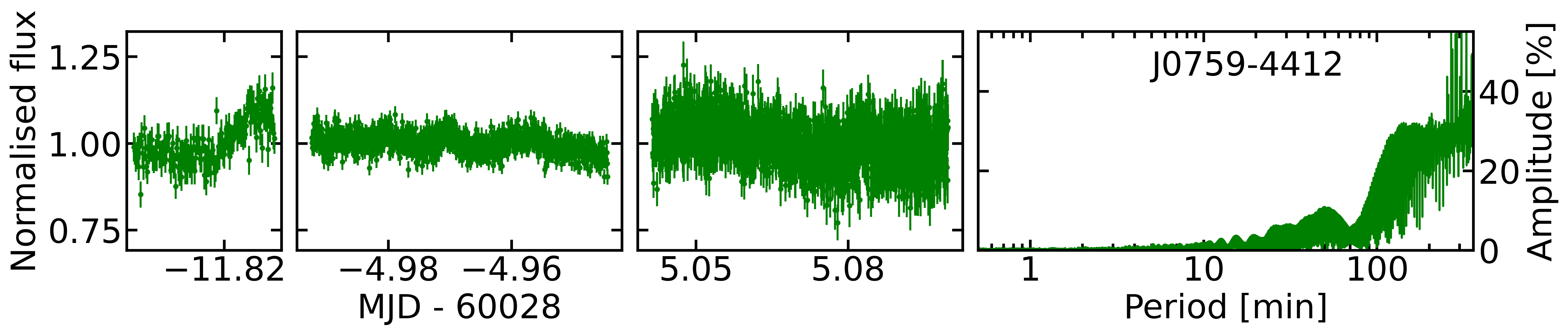}
    \includegraphics[width=\columnwidth]{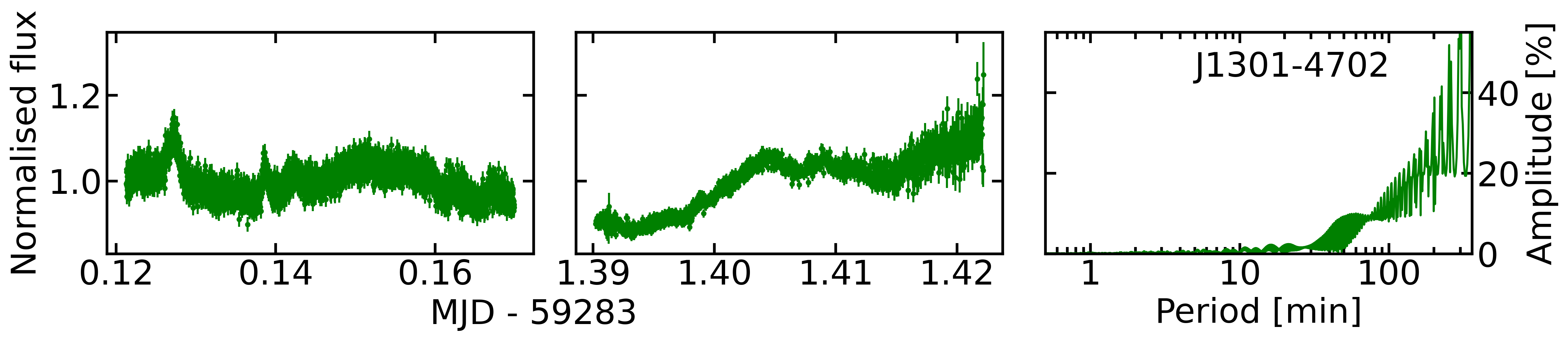}
    \includegraphics[width=\columnwidth]{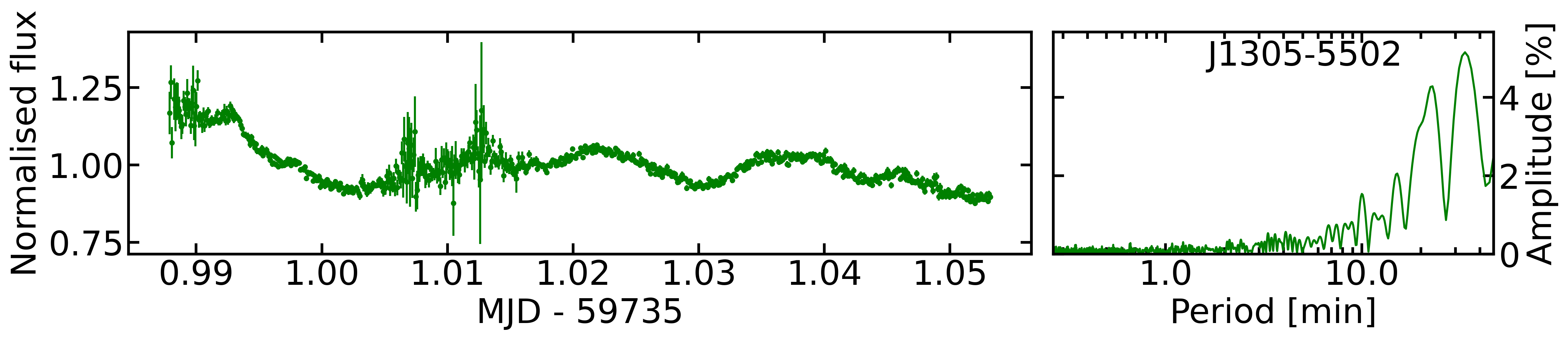}
    \includegraphics[width=\columnwidth]{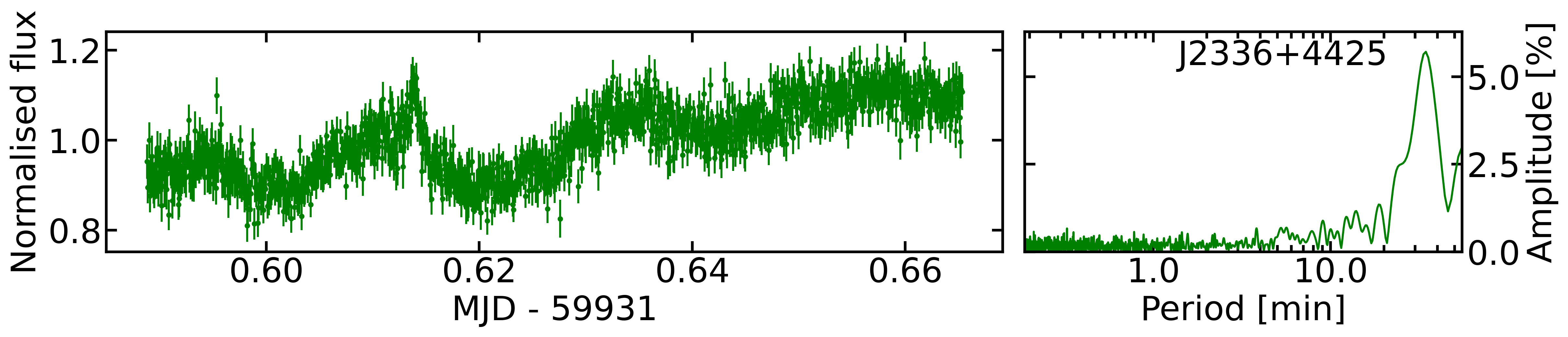}
    \caption{The $g$-band light curves (left panels) and periodogram (rightmost panel) for the six systems showing low-level aperiodic variability and flaring. For J0657-0615, which appeared constant on the second night, only the first night is included in the Fourier transform.}
    \label{fig:ucam_ysos}
\end{figure}

Another six systems showed low-level variability and/or flaring, consistent with the behaviour of YSOs. These are shown in Fig.~\ref{fig:ucam_ysos}. The other nine systems observed with ULTRACAM/ULTRASPEC showed no detectable variability in the optical data. As they were selected based on infrared variability, the lack of optical variability might be because variability is concentrated in the infrared, or because the infrared variability is spurious due to background contamination.

\subsection{Spectroscopy}

Time-series photometry alone can be insufficient to provide a secure classification, as different types of systems can show similar variability characteristics. Therefore, we also strove to obtain spectroscopy for previously uncharacterised systems. 

% section on X-shooter data

\subsubsection{XSHOOTER at the VLT}

The four previously uncharacterised systems that had reported X-ray detections (J0007-6759, J0156-8358, J0354-1652, J0435-7527) and additionally two systems showing UV detection and periodic variability in photometric surveys (J0428-3300, J1912-4410) were observed with XSHOOTER at ESO's 8.2-m Very Large Telescope (VLT; proposals 108.228J and 109.234F). Two spectra on consecutive nights were obtained for all targets (except J0428-3300, which was not observed a second time before the program was terminated) to probe for spectral variability and/or radial velocity changes. The spectra were automatically reduced by the XSHOOTER pipeline.

\begin{figure}
    \includegraphics[width=\columnwidth]{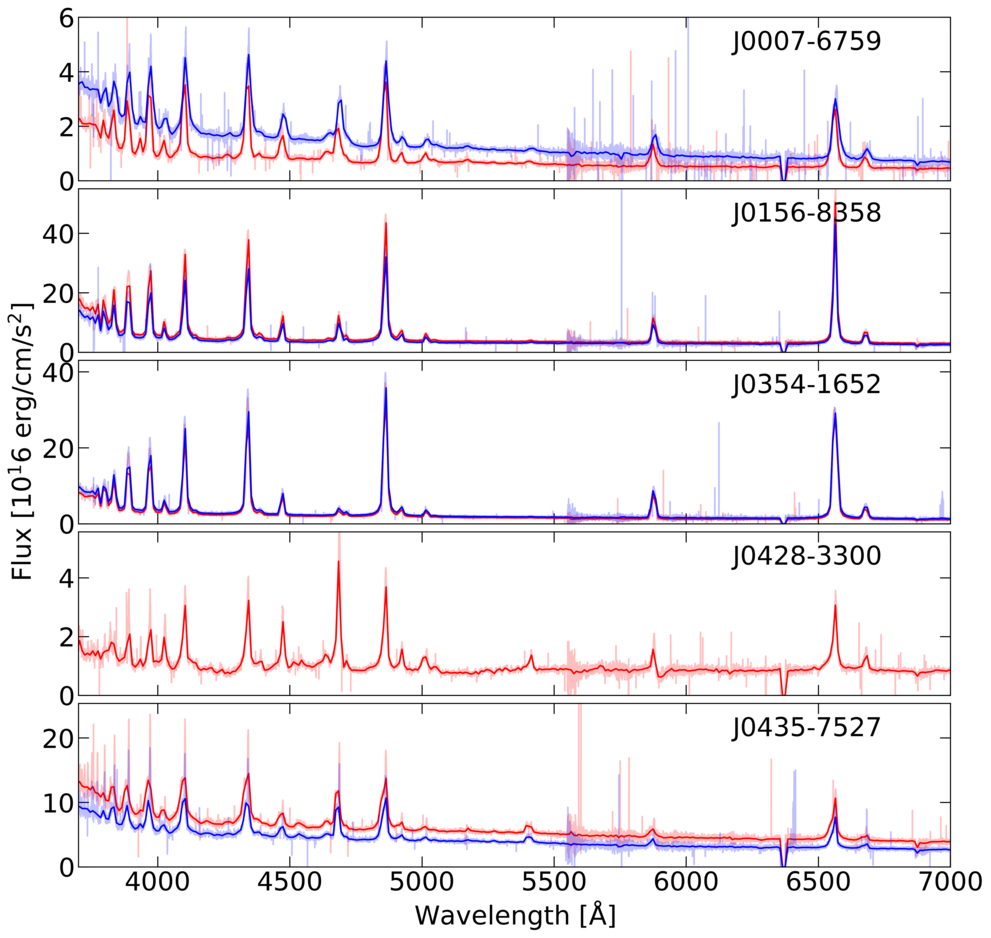}
    \caption{XSHOOTER spectra for J0007-6759, J0156-8358, J0354-1652, J0428-3300 and J0435-7527 (from top to bottom). The transparent lines show the actual spectra and the solid lines an average every 50 points. The first spectrum obtained is shown in red and the second (when available) in blue.}
    \label{fig:xshooter}
\end{figure}

J1912-4410 showed a spectrum with a blue and a red component, with very narrow lines indicating absence of accretion, like seen for AR~Sco, and its analysis was published in \citet{Pelisoli2023}. The spectra for the other systems are shown in Fig.~\ref{fig:xshooter}. All except J0156-8358 and J0354-1652 show strong HeII emission (with strength comparable to the Balmer lines), which is a characteristic of magnetic systems with channelled accretion \citep[e.g.][]{Voikhanskaya1987}. J0007-6759, J0428-3300, and J0435-7527 show asymmetric line profiles that change from night to night (when more than one night was available). This suggests that they are polars -- the asymmetry is caused by the combination of the narrow emission component from the bright spot with the broader emission from the accretion stream and irradiated companion. J0156-8358 and J0354-1652, on the other hand, show symmetric emission lines and are more likely non-magnetic CVs, given the weak HeII emission.

% section on other data

\subsubsection{OSIRIS at the GTC}
\label{sec:GTC}

Spectroscopic observations of J0452+3017 were performed with the 10.4\,m Gran Telescopio Canarias (GTC) at the Observatorio del Roque de los Muchachos. The object was observed in service mode in a 2.5-hour observing block on 2023 March 16, covering two cycles of the period detected in archival data. The spectra were taken with the Optical System for Imaging and low-intermediate-resolution Integrated Spectroscopy (OSIRIS) spectrograph \citep{osiris}, which consists of a mosaic of two Marconi CCDs, each with $2048\times4096$ pixels. Each pixel has a physical size of 15\,$\mu$m. We used $1\times1$ binning for our observations, giving a plate scale of 0.127\,arcsec. OSIRIS was used in long-slit mode, centering the object in CCD2. We used the R500R grism, providing $4800 - 10000$\,\AA\ coverage at $R=587$ resolution. 21 spectra of the object were obtained with 3600-sec exposure times, in dark, spectroscopic conditions with  1.3-1.4'' seeing; the slit width was set to 1.0''. The data were reduced using {\sc iraf} to perform bias subtraction, flat-field correction and spectral extraction.

\begin{figure}
    \includegraphics[width=\columnwidth]{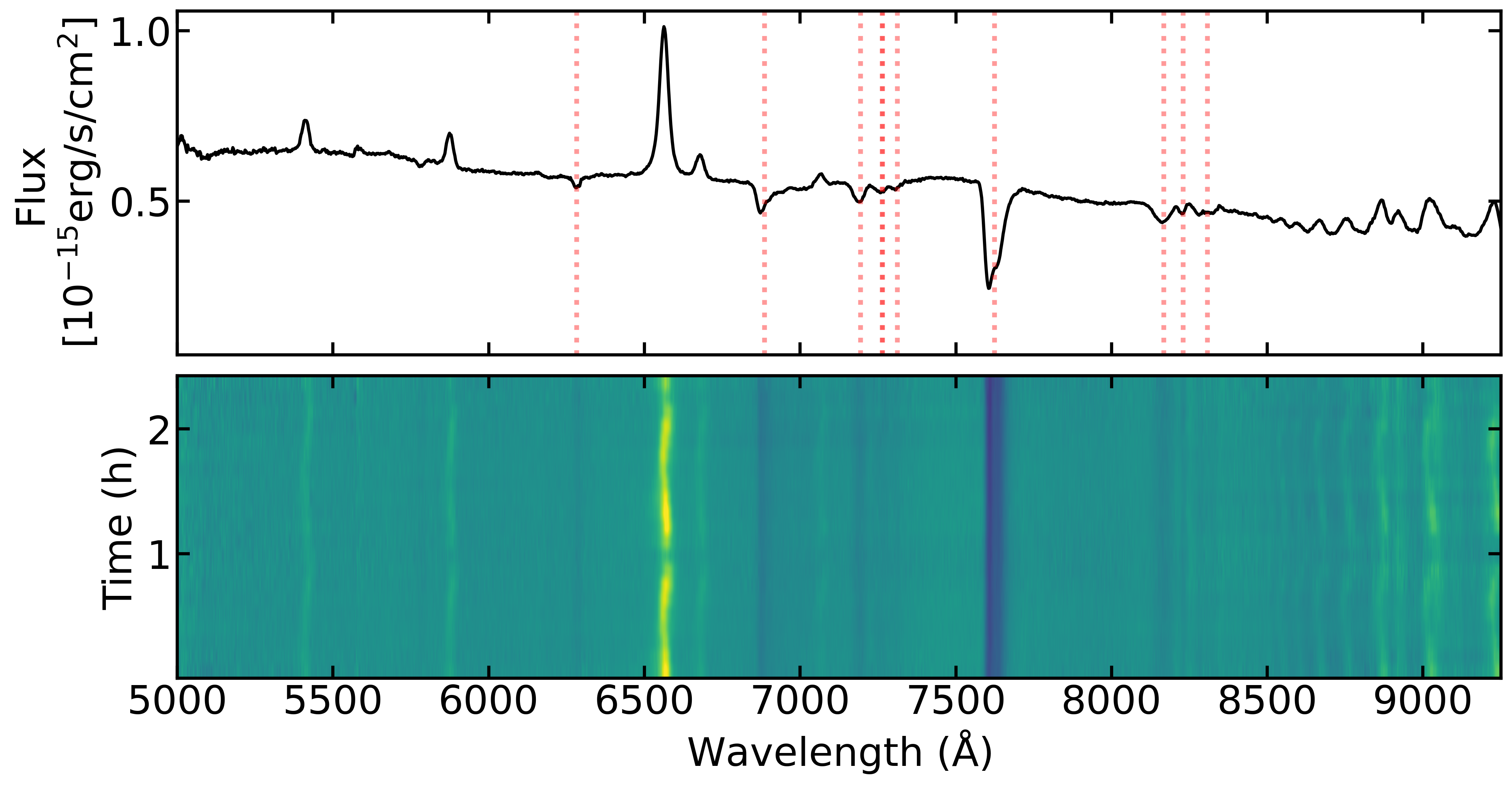}
    \caption{The top panel shows the co-added spectrum of J0452+3017 (telluric absorption lines are indicated by red dashed lined). The bottom panel shows the trailed spectra, which show that the radial velocity of the emission lines is changing periodically.}
    \label{fig:J0452_spec}
\end{figure}

Fig.~\ref{fig:J0452_spec} shows the reduced co-added spectrum of J0452+3017 as well as the trailed spectra. The spectrum shows broad emission components, in particular H$\alpha$, whose radial velocity can be seen to vary periodically in the trail, a behaviour consistent with an IP. We estimated the radial velocity of the most prominent emission lines (H$\alpha$, HeI 5875~\AA, HeII 5411~\AA) by fitting Gaussian profiles with a fixed width (determined from the co-added spectrum) and variable centre. The radial velocities were fit to obtain the orbital ephemeris with the period fixed to the one derived from the TESS data ($1.325\pm0.008$~h). We obtained
\begin{eqnarray}
BJD(TDB) = 2460020.43323(69) + 0.0552(3)E
\label{eq:j0452_eph}
\end{eqnarray}
where BJD(TDB) is the barycentric Julian date in the Barycentric Dynamical Time (TDB) scale, and $E$ is an integral cycle number. The obtained radial velocity amplitude was found to be $170.9\pm16.0$~km/s, implying a binary mass function of $0.1029\pm0.0252$~M$_{\odot}$. The emission lines likely come from the irradiated surface of the companion and, as a consequence, the derived radial velocities are not those of the centre of mass of the companion. In fact, considerable scatter can be seen, suggesting that the different lines might originate on different regions on the surface of the companion, depending on the temperature distribution. Therefore, the derived radial velocity amplitude and mass function are lower limits for this system.

% Radial velocity fit
%T0 = 60019.93323 +/-     0.00069 days
%V = 170.9 +/- 16.0 km/s
%K = 259.7 +/- 21.1 km/s
%Mass function = 0.1029 +/- 0.0252 MSun

% plot of spectrum showing that there's no contribution from companion, together with trailed spectra

% plot of light curve and RVs on the same ephemeris

\begin{figure}
    \includegraphics[width=\columnwidth]{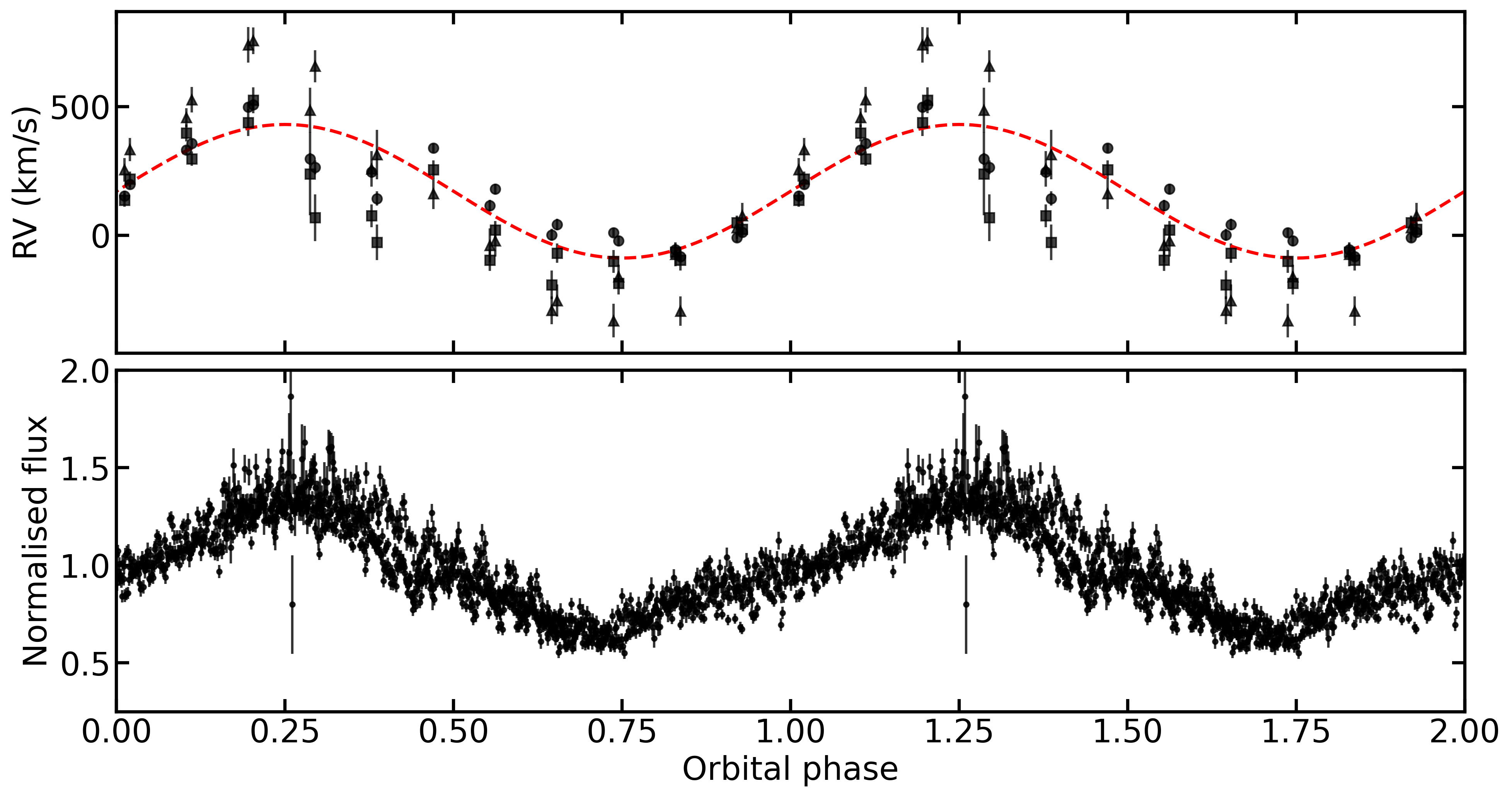}
    \caption{The radial velocities (top) and ULTRASPEC photometry (bottom) for J0452+3017 folded to the same orbital ephemeris given in Equation~\ref{eq:j0452_eph}. The red dashed line in the top panel shows the fit to the radial velocities.}
    \label{fig:J0452_phase}
\end{figure}

\subsection{Other instruments}

\begin{figure}
    \includegraphics[width=\columnwidth]{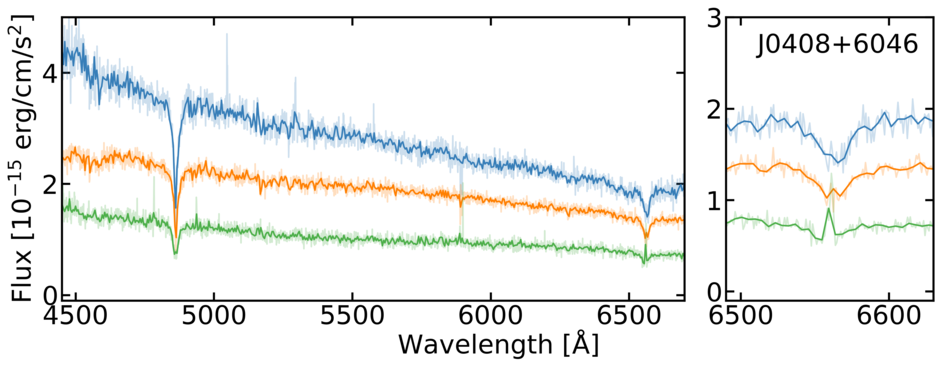}
    \includegraphics[width=\columnwidth]{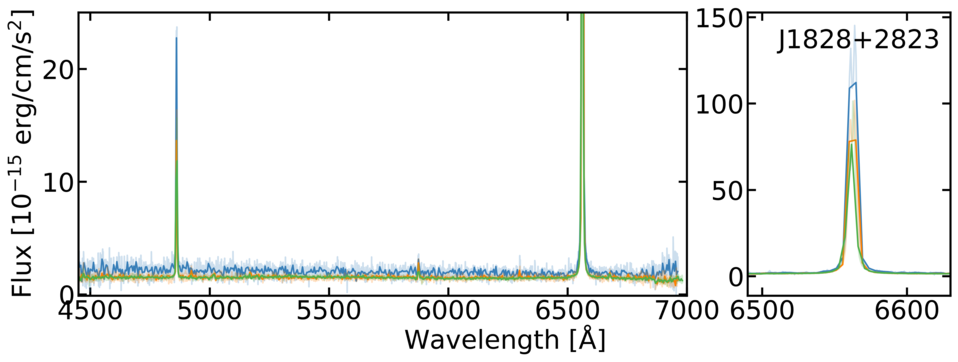}
    \caption{Spectra for J0408+6046 (top) and J1828+2823 (bottom) obtained with INT/IDS. }
    \label{fig:int_spec}
\end{figure}

J0408+6046 and J1828+2823 were observed with the Intermediate Dispersion Spectrograph (IDS) at the Isaac Newton Telescope (INT) as part of a search for CVs unrelated to this work. J0408+6046 was observed on three nights (2020 February 12, 2020 August 20, and 2020 December 17) and J1828+2823 on four nights (2020 May 15, 2020 September 10, 2020 September 12, and 2021 June 04), in all cases using the R632V grating. The spectra were reduced with {\sc pamela} and {\sc molly} \citep{Marsh1989} and are shown in Fig.~\ref{fig:int_spec}. J0408+6046 shows a spectrum typical of a pre-main sequence Herbig Ae star \citep{Herbig1960}, with narrow Balmer lines and variable emission (most notably seen in H$\alpha$). J1828+2823 shows strong and narrow Balmer emission lines with no radial velocity variability, consistent with a YSO.

\begin{figure}
    \includegraphics[width=\columnwidth]{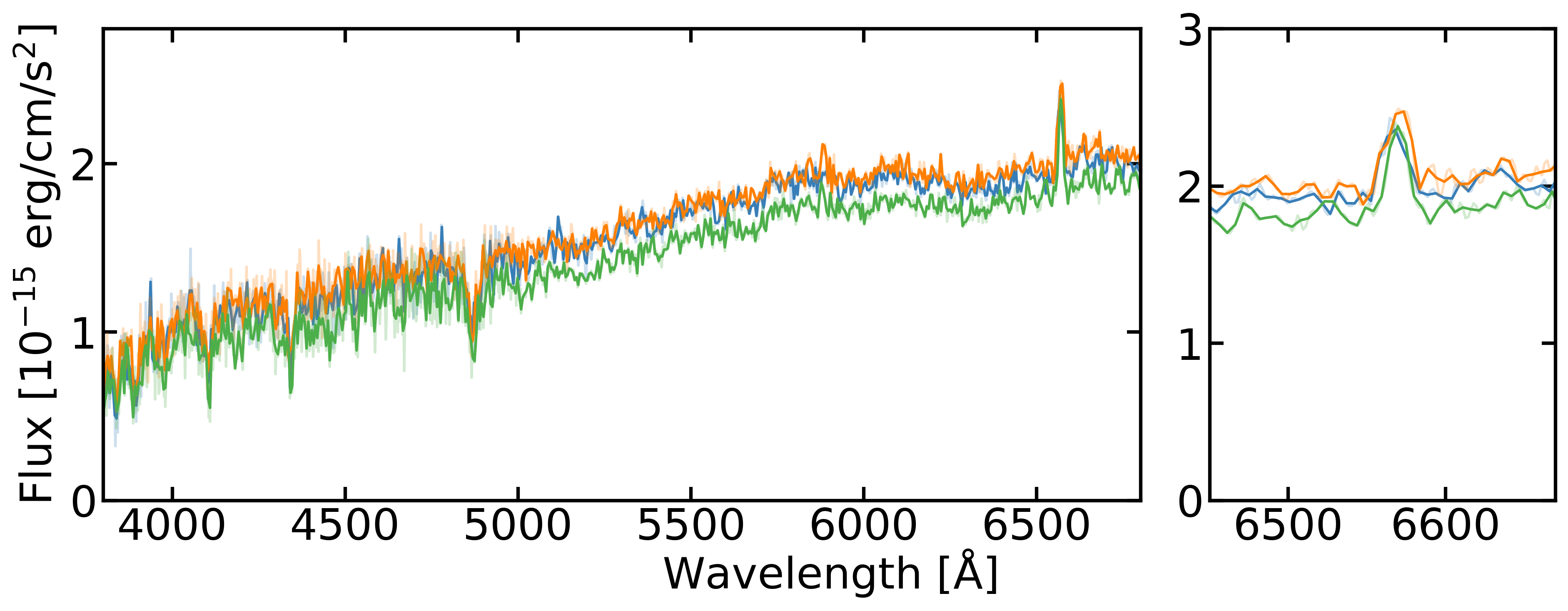}
    \caption{FLOYDS spectra of J1917-1231.}
    \label{fig:floyds_spec}
\end{figure}

J1917-1231 was observed as backup target with FLOYDS during a follow-up survey of AM CVn systems. The pipeline-reduced spectra are shown in Fig.~\ref{fig:floyds_spec}. Similar to J0408+6046, it shows features consistent with a Herbig Ae star.

\begin{figure}
    \includegraphics[width=\columnwidth]{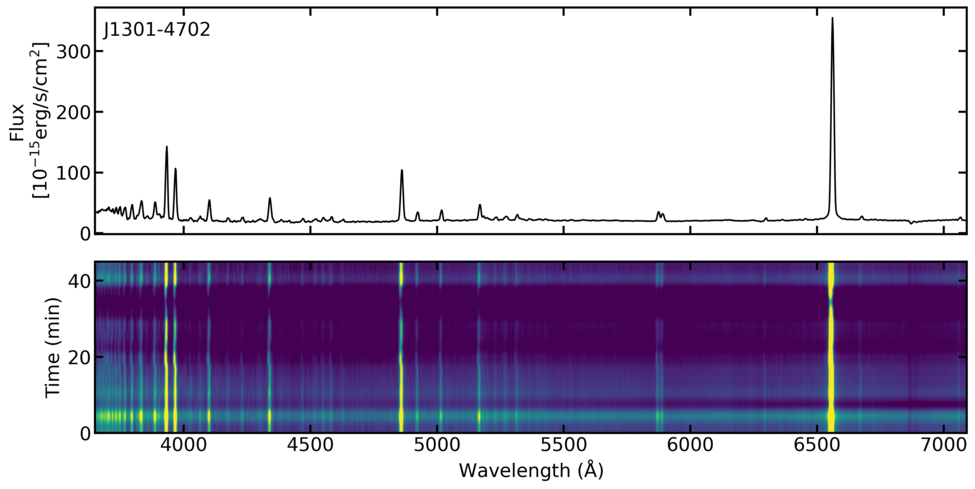}
    \includegraphics[width=\columnwidth]{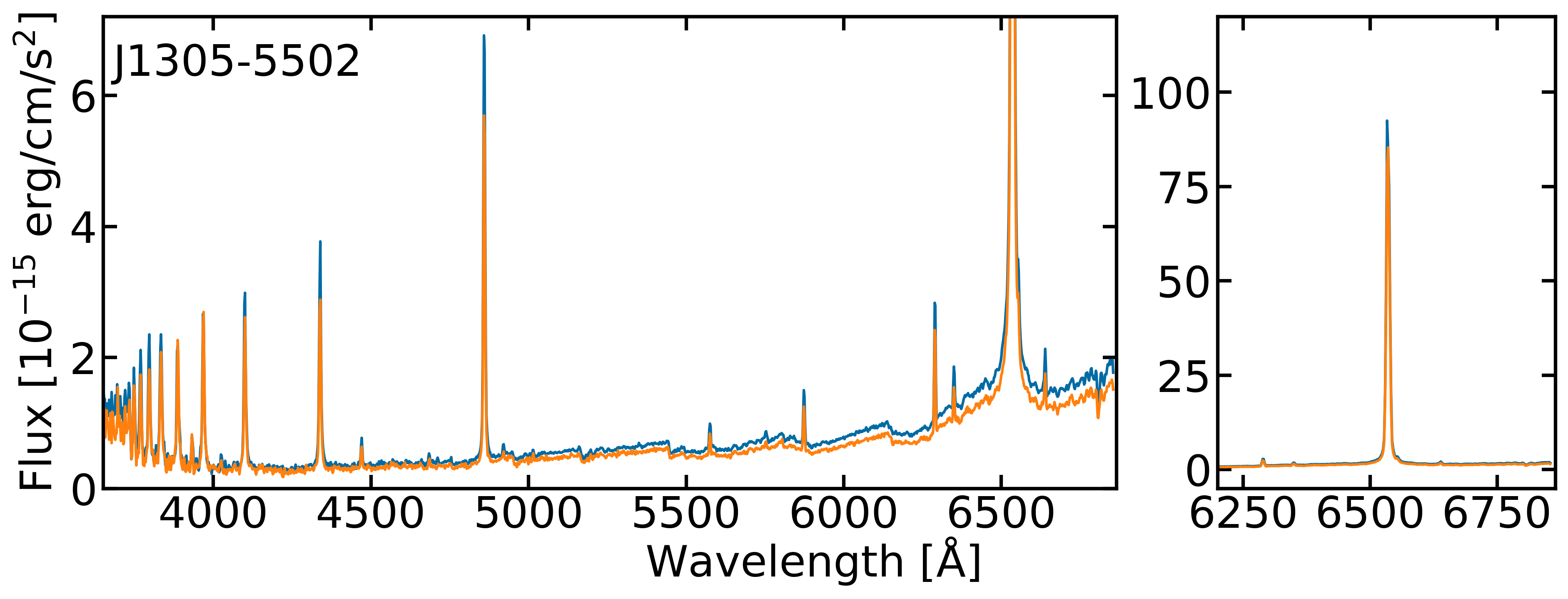}
    \caption{Co-added (top) and trailed (middle) spectra for J1301-4702, and the two spectra of J1305-5502, obtained with SOAR.}
    \label{fig:soar_spec}
\end{figure}

J1301-4702 and J1305-5502 were observed as backup targets with Goodman at SOAR using the 400~l/mm grating. Sixteen consecutive spectra were obtained for J1301-4702, covering a timespan of about 40~min. For J1305-5502, two consecutive spectra were obtained. The spectra were reduced with {\sc iraf} and are shown in Fig.~\ref{fig:soar_spec}. In both cases, narrow emission lines with no radial velocity variability are seen, suggesting these are also YSOs.

\section{Results and Discussion}

\subsection{Young stellar objects}

Our follow-up observations suggest that ten out of the 26 previously uncharacterised objects are consistent with YSOs. J0408+6046 and J1917-1231 can be classified as Herbig Ae stars based on their spectra (Fig.~\ref{fig:int_spec}, top and Fig.~\ref{fig:floyds_spec}, respectively). Similarly, the narrow lines with no radial velocity variability seen for J1828+2823 (Fig.~\ref{fig:int_spec}, bottom), J1301-4702 and J1305-5502 (both shown in Fig.~\ref{fig:soar_spec}) are consistent with T~Tauri stars. Another system that is likely a YSO is J0539+0824, which showed variability with no determined period in archival data (Fig.~\ref{fig:ZTF_aperiodic}), but no clear variability when observed with high-speed photometry, consistent with slow stochastic variability from a circumstellar disc. Additionally, it has been identified as potential member of a stellar association \citep{Spina2021}, which supports the interpretation as a YSO. J0551+0752, J0657-0615, J0759-4412, and J2336+4425 show a similar variability behaviour to J0539+0824, with only low-level variability and/or flaring (as shown in Fig.~\ref{fig:ucam_ysos}), and thus our interpretation is that they are also YSOs. 

Six out of these ten possible YSOs are in the catalogue of \citet{Marton2019}, which used machine learning methods to identify YSO candidates from {\it Gaia} DR2 and WISE data. J0657-0615 appears as a high-probability (0.96) YSO, and is also classified as such by \citet{Wilson2023}. J1305-5502 also has a moderately high YSO probability (0.85, when the WISE W3 and W4 bands are not included). The four other targets have a higher probability of being extragalactic objects according to their classification: J0408+6046 (0.64), J0539+0824 (0.98), J0551+0752 (0.79), and J0759-4412 (0.95). We can rule out an extragalactic origin for J0408+6046 based on its Herbig Ae-type spectrum, but the possibility of extragalactic origin remains open for the other three objects.

\subsection{Polars}

Three systems showed spectra (see Fig~\ref{fig:xshooter}) and light curves consistent with a polar: J0007-6959, J0428-3300 and J0435-7527. They are all newly discovered polars; J0007-6959 was classified as a white dwarf candidate by \citet{GentileFusillo2019}. Curiously, J0435-7527 was classified as an RR~Lyra based on its CRTS light curve \citep{Drake2017}. The orbital periods for these systems were determined from the TESS data (see Table~\ref{tab:photometry} and Fig.~\ref{fig:tess_periodic}) and are $1.5826\pm0.0040$, $2.5466\pm0.0011$, and $2.36\pm0.23$~h, placing two of them within the CV period gap between 2-3~hours, believed to occur when the donor star becomes fully convective and magnetic braking becomes inefficient, giving the binary time to detach. This is in line with findings that polars do not actually show a period gap \citep[e.g][]{Schreiber2024}.

J0428-3300 and J0435-7527 additionally show evidence for quasi-periodic oscillations (QPOs), as illustrated in Fig.~\ref{fig:ucam_qpos}. These are variations that show coherence over a limited amount of time and are a common feature of polars \citep[e.g][]{Bera2018}, having been detected from infrared to X-rays. They generally group in two ranges of periods, a few (1--5) seconds or a few (4--10) minutes, with the latter being the most common feature in the optical \citep{Potter2010}. The QPOs observed for J0428-3300 and J0435-7527 at $3.676\pm0.009$ and $4.24\pm0.11$~min, respectively, are therefore consistent with a polar behaviour. The occurrence of QPOs, and other photometric variations seen at polars such as flickering, are believed to be caused by variations in the accretion flow, possibly due to inhomogeneities \citep{Kuijpers1982}, modulation at the edge of the white dwarf's magnetosphere \citep{Patterson1981}, or modulation near the L1 point due to irradiation \citep{King1989}.

\subsection{Intermediate polar}

J0452+3017 showed an intermittent pulsing behaviour in its ULTRASPEC light curve with a period of $3.0\pm0.5$~min (Fig.~\ref{fig:ucam_j0452}), that prompted further follow-up. The obtained GTC spectra revealed broad and symmetric emission (Fig.~\ref{fig:J0452_spec}) which, combined with the detection of a short spin period, suggest that this system is an intermediate polar. We have found its orbital period to be $1.325\pm0.008$~h, which makes it, to our knowledge, the intermediate polar with the shortest known orbital period \footnote{See https://asd.gsfc.nasa.gov/Koji.Mukai/iphome/catalog/alpha.html}. 

\subsection{Likely non-magnetic CVs}

The spectra of J0156-8358 and J0354-1652 (in Fig.~\ref{fig:xshooter}) show only weak HeII emission, suggesting they are non-magnetic CVs. Their light curves (Fig.~\ref{fig:ucam_stochastic}), showing only stochastic variability on top of the orbital period -- that is, with no clear detection of a spin period that could indicate an intermediate polar nature -- support that conclusion. We put J1425-6837 in this category as well given the observed sudden change in brightness state (shown in Fig.~\ref{fig:ucam_j1425}), which is consistent with the behaviour of a dwarf nova, although some intermediate polars have been observed to show outbursts. 

J0156-8358 appears in the literature only as an X-ray source, therefore its nature is determined for the first time here. Its orbital period is $2.348\pm0.035$~h, which interestingly places it in the period gap. The depth of this gap depends somewhat on the observed sample \citep[see e.g.][]{Inight2023}, therefore finding systems within the gap is not completely unexpected; nevertheless, this makes this system a valuable addition to the CV population given the relative rareness of period-gap systems.

J0354-1652, on the other hand, was studied by \citet{Thorstensen2006} and more recently by \citet{Joshi2022}. \citet{Thorstensen2006} found a hint of radial velocity variability at a period of 46 min and had no clear conclusions on the nature of the system, which is why we kept in our sample for further follow-up. Like \citet{Joshi2022}, we found no indication of a 46~min period in the data. They report a likely orbital period of $1.689\pm0.001$~h, consistent with our findings. They similarly notice the weakness of HeII as a sign of lack of magnetism, though they remark that the variability is akin to that seen in polars. We favour a non-magnetic classification based on the spectral features.

\subsection{Other systems}

J1125+5012 is in the vicinity ($< 7$") of a bright ($G= 13.2$) source, hence its photometry is likely contaminated by diffraction spikes from this neighbour. As a consequence, its selection as a candidate is unreliable and we pursued no follow-up, leaving its nature as undetermined.

Seven systems (J0808-0935, J1226-2304, J1657-2631, J1845-2146, J2059-0747, J2101-1616 and J2119-2308) showed no optical variability in either photometric surveys or high-speed follow-up, which ruled out a nature as binary white dwarf pulsars. Their nature is unclear based on existing data alone and we cannot rule out that the infrared variability that led to their selection is caused by background sources.

\subsection{J191213.72-441045.1}
\label{sec:j1912}

The main discovery of this targeted search was J1912-4410, reported in detail in \citet{Pelisoli2023}. It was selected as a candidate as its position in the {\it Gaia} colour-magnitude diagram, $W1-W2$ colour (0.685) and WISE variability flag ({\tt 999n}) all fit the selection criteria. Subsequent follow-up found it to show pulsing behaviour in the optical with a period of 5.3~min, interpreted as the spin of the white dwarf. Pulses were also detected in the radio and in X-rays \citep{Schwope23}. Like in AR~Sco, the white dwarf has an M-type irradiated companion, and the orbital period determined from the TESS data (shown here in Fig~\ref{fig:tess_periodic}) is 4.03~h. The spectra of the system show narrow emission lines consistent with originating on the surface of the irradiated companion suggesting no persistent accretion, with the pulsed emission likely originating due to magnetic interaction between the two stars \citep[][]{Pelisoli2024a}. All of these characteristics led to its classification as a second binary white dwarf pulsar.

\section{Summary and Conclusions}

We carried out a targeted search for binary white dwarf pulsars combining the astrometric and photometric data from {\it Gaia} with infrared data and variability flags from WISE. Based on criteria determined using the only previously known binary white dwarf pulsar, AR~Sco, we identified 56 candidates (Table~\ref{tab:candidates}), 26 of which were previously uncharacterised systems. The remaining 30 were known in the literature and a mix of CVs and YSOs. We analysed archival times-series photometry for all previously unknown systems and carried out high-speed photometry and spectroscopy follow-up observations, prioritising systems with detected variability.

\begin{figure*}
    \includegraphics[width=0.8\textwidth]{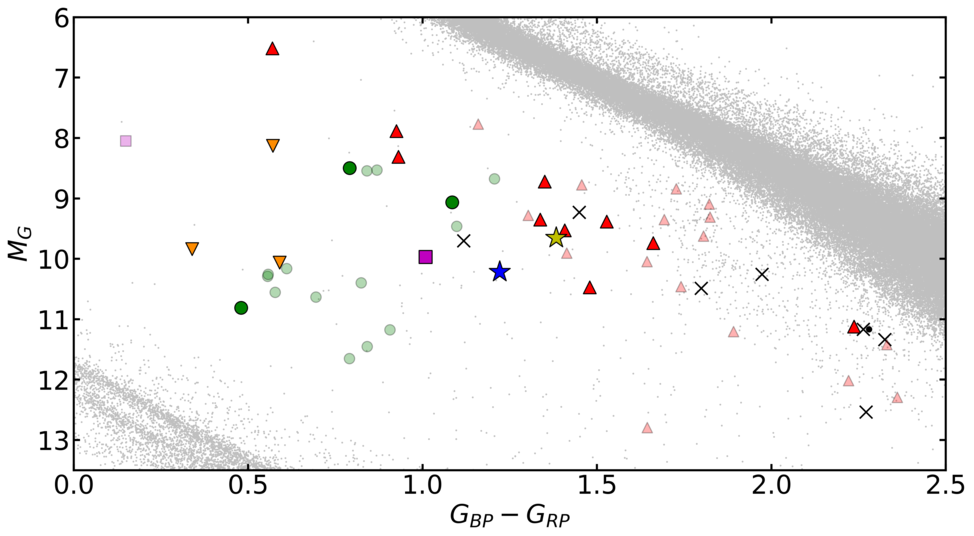}
    \caption{The location in the {\it Gaia} colour-magnitude diagram of the 56 binary white dwarf pulsar candidates. AR~Sco is the yellow star and J1912-4410, discovered as a result of this work, is the blue star. Polars are shown as green circles, one IPs as a magenta squares, and red triangles are YSOs. The symbols are smaller and fainter for previously known objects. The orange upside-down triangles are the likely non-magnetic CVs, the black dot is the blended object that was not further followed up, and the black crosses showed no detectable variability in follow-up optical observations.}
    \label{fig:candidates}
\end{figure*}

We concluded that ten of the followed-up objects are YSOs, three are polars, one is an intermediate polar, and three are likely non-magnetic CVs. One system was not followed up due to blending, and seven systems showed no optical variability in either archival or follow-up light curves and were not further studied. Finally, one new binary white dwarf pulsar, J1912-4410, was identified. The location of these sources in the {\it Gaia} colour-magnitude diagram is shown in Fig.~\ref{fig:candidates}. The YSOs concentrate near the main sequence as expected, with the identified CVs (magnetic or otherwise) being located closer to the white dwarf cooling sequence. Like AR~Sco, J1912-4410 is located somewhat between these two regions. A detailed characterisation of this system was presented in a dedicated work \citep{Pelisoli2023}.

Although our search successfully identified one new binary white dwarf pulsar, that came at the expense of a large number of contaminant sources, in particular YSOs. We believe that the YSO contamination is mainly driven by our use of infrared data, especially variability, as a criterion, given that YSOs are infrared bright and typically variable. Curiously, although we were fairly unrestricted in the colour-magnitude magnitude selection, essentially targeting anything between the main sequence and white dwarf cooling track, J1912-4410 was found in a position very close to AR~Sco. This might suggest that a narrow range of orbital and stellar properties, resulting in these observed photometric properties, is required to trigger the pulsed radio emission observed for J1912-4410 and AR~Sco. Future searches might benefit from focussing on this region and relaxing or removing infrared criteria.

\section*{Acknowledgements}

IP acknowledges support from the Royal Society through a University Research Fellowship (URF\textbackslash R1\textbackslash 231496).
VSD and ULTRACAM/ULTRASPEC operations are funded by the Science and Technology Facilities Council (grant ST/Z000033/1). GT was supported by grant IN109723  from the  \textit{Programa de Apoyo a Proyectos de Investigación e Innovación Tecnológica} (PAPIIT).  Part of the data was based on observations made with the GTC telescope, in the Spanish Observatorio del Roque de los Muchachos of the Instituto de Astrofísica de Canarias, under Director’s Discretionary Time. This work was supported by Naresuan University (NU), and National Science Research and Innovation Fund (NSRF), Grant No. R2567B015. This work has made use of data obtained at the Thai National Observatory on Doi Inthanon, operated by NARIT. This project has received funding from the European Research Council (ERC) under the European Union’s Horizon 2020 research and innovation programme (Grant agreement No. 101020057).

%%%%%%%%%%%%%%%%%%%%%%%%%%%%%%%%%%%%%%%%%%%%%%%%%%

\section*{Data Availability}

The data analysed in this work can be made available upon reasonable request to the authors.

%%%%%%%%%%%%%%%%%%%% REFERENCES %%%%%%%%%%%%%%%%%%

% The best way to enter references is to use BibTeX:

\bibliographystyle{mnras}
\bibliography{arscos}

%%%%%%%%%%%%%%%%%%%%%%%%%%%%%%%%%%%%%%%%%%%%%%%%%%

%%%%%%%%%%%%%%%%% APPENDICES %%%%%%%%%%%%%%%%%%%%%

\appendix

\section{Previously known systems}
\label{apx:known}

\subsection{Polars}

\noindent \textbf{V* BL Hyi:} a well-studied polar with a period of 100~min \citep[e.g.][]{Visvanathan1982}.

\noindent \textbf{CRTS J035758.7+102943:} a polar with a period of 1.900344(24)~h \citep{Schwope2012,Thorstensen2016}.

\noindent \textbf{V* UW Pic:} a known polar with an orbital period of 2.225~hours and a magnetic field strength of 19~MG \citep{Reinsch1994}.

\noindent \textbf{PPMXL 3189652360374206258:} classified as a polar by \citet{Halpern2015}, who report a period of 1.70177(26)~h. 

\noindent \textbf{V* HS Cam:} known eclipsing polar with a period of 1.637~h \citep{Tovmassian1997}.

\noindent \textbf{V* VV Pup:} a polar \citep{Tapia1977} showing quasi periodic oscillations \citep{Bonnet2020}. It has an orbital period of 1.67~h \citep{Walker1965} and a magnetic field structure that is well-described by a 40~MG offset dipole \citep{Mason2007}. 

\noindent \textbf{V* V834 Cen:} a well-known polar showing quasi-periodic oscillations \citep{Mouchet2017}. The orbital period is 1.692~h \citep{Mason1983} and it has a magnetic field of 23~MG \citep{Ferrario1992}.

\noindent \textbf{2MASS J14534105-5521387:} this is a polar showing quasi-periodic oscillations. Its orbital period is 3.1564(1)~h, and it was characterised by \citet{Potter2010}.

\noindent \textbf{V* V2301 Oph:} a well-characterised eclipsing polar also known as H 1752+081. The orbital period is 1.882801~h \citep{Barwig1994, Silber1994}. It has a relatively low magnetic field strength for a polar, of 7~MG \citep{Ferrario1995}.

\noindent \textbf{V* EP Dra:} also known as H~1907+690, this is a well characterised eclipsing polar. The reported orbital period is 1.743750~h \citep{Remillard1991}, and the magnetic field strength is 16 ~MG \citep{Schwope1997}. 

\noindent \textbf{2MASS~J19293303-5603434:} an eclipsing polar with an orbital period of $92.5094\pm0.0002$~min recently discovered by \citet{Schwope2022}. 

\noindent \textbf{Gaia DR2 1956566510538468224:} there was a {\it Gaia} alert for these system, Gaia18aya. It was followed-up by \citet{Thorstensen2020} who identified it as a polar with magnetic field strength of $\sim 50$~MG and orbital period of $2.002757(38)\pm$~hours.

\noindent \textbf{RX J2218.5+1925:} also known as Swift~J2218.4+1925, this is a known polar with an orbital period of of $2.158\pm0.003$~hours. It was initially classified from optical spectroscopy by \citet{Thorstensen2009}, and further characterised from X-ray data by \citet{Bernardini2014}.

\subsection{Intermediate polar}

\noindent \textbf{PPMXL 2431202680279124324:} an intermediate polar with a spin period of 1412~s and orbital period of 81~min \citep{Halpern2015}.

\subsection{Young stellar objects}

\noindent \textbf{SSS~J035055.8-204817:} \citet{Pala2020} obtained spectra of this object and concluded it to be a YSO due to the presence of lithium and of forbidden lines of [OI] (5577 and 6300~\AA) and [S II] (6730~\AA), commonly observed for YSOs but not CVs.

\noindent \textbf{2MASS J05301240+0148214:} a T~Tauri star in the Orion OB1 association \citep{Briceno2019}.

\noindent \textbf{2MASS J05315396+0242310:} reported as a likely classical T-Tauri star by \citet{vanEyken}.

\noindent \textbf{2MASS J05371640-0711463:} a young stellar object in the Orion A molecular cloud \citep{Grosschedl2019}.

\noindent \textbf{Gaia DR2 3336335474617325568:} a member of the Collinder 69 open cluster \citep{Cantat2018}, which is $\approx 12$~Myr old \citep{Cantat2020}.

\noindent \textbf{2MASS J05422602-0308566 and 2MASS J05405519-0247497:} H$\alpha$ emission-line stars in a molecular cloud towards the M42 open cluster \citep{Pettersson2014}.

\noindent \textbf{Gaia DR2 5329108769930689664:} is a member of the open cluster IC 2395 \citep{Cantat2018}, which is $\approx 20$~Myr old \citep{Cantat2020}.

\noindent \textbf{[FLG2003] eps Cha 11:} this a pre-main sequence star in the $\eta$ Chamaeleontis cluster that is viewed through an edge-on disc \citep{Lyo2008}.

\noindent \textbf{BPS CS 30311-0012:} this is classified as a young (sub)stellar object by \citet{Gagne2015}. 

\noindent \textbf{CRTS J155929.1-223618 and EPIC 204388640:} these are young stars in the direction, and possible members, of the Upper Scorpius association \citep{Luhman2020}.

\noindent \textbf{2MASS J16063093-2258150:} this object has been spectroscopic classified as a galaxy by \citet{Luhman2018}, which would require its {\it Gaia} parallax of $7.270\pm0.220$~mas to be very inaccurate. More recently, relying on {\it Gaia} data, it was classified as a candidate member of the Upper Scorpius association \citep{Luhman2020}.

\noindent \textbf{Gaia DR2 6048855674228736128:} has been identified as a candidate member of the association $\rho$ Ophiuchi by \citet{Canovas2019}.

\noindent \textbf{2MASS J18294021+0015127:} a well-characterised young stellar object in the Serpens association \citep{Harvey2007,Oliveira2009}.

\noindent \textbf{2MASS J21205785+6848183:} this is a pre-main sequence star that is part of the G109+11 association, as determined by \citet{Kun2009}. They remark that the precise nature of the system is uncertain, with observed properties being consistent with an edge-on disk or an unresolved binary.

%%%%%%%%%%%%%%%%%%%%%%%%%%%%%%%%%%%%%%%%%%%%%%%%%%

% Don't change these lines
\bsp	% typesetting comment
\label{lastpage}
\end{document}